\documentclass[aps, prr, amsmath,amssymb,aps,floatfix, twocolumn, superscriptaddress]{revtex4-2}

\usepackage[utf8]{inputenc}
\usepackage{graphicx}
\usepackage{dcolumn}
\usepackage{bm}
\usepackage{color}
\usepackage[normalem]{ulem}

\def\spose#1{\hbox to 0pt{#1\hss}}
\def\lesssim{\mathrel{\spose{\lower 3pt\hbox{$\mathchar"218$}}
 \raise 2.0pt\hbox{$\mathchar"13C$}}}
\def\gtrsim{\mathrel{\spose{\lower 3pt\hbox{$\mathchar"218$}}
 \raise 2.0pt\hbox{$\mathchar"13E$}}}
\definecolor{darkgreen}{rgb}{0.0, 0.5, 0.0}

\begin{document}
\title{Generative Modeling of Entangled Polymers with a Distance-Based Variational Autoencoder}
\author{Pietro Chiarantoni}
\affiliation{Institute for Computational Molecular Science, Temple University, Philadelphia, PA, 19122}
\affiliation{Institute for Genomics and Evolutionary Medicine, Temple University, Philadelphia, PA, 19122}
\affiliation{Temple Materials Institute, Temple University, Philadelphia, PA, 19122}
\affiliation{Department of Biology, Temple University, Philadelphia, PA, 19122}

\author{Oscar Serra}
\affiliation{Institute for Computational Molecular Science, Temple University, Philadelphia, PA, 19122}
\affiliation{Institute for Genomics and Evolutionary Medicine, Temple University, Philadelphia, PA, 19122}
\affiliation{Temple Materials Institute, Temple University, Philadelphia, PA, 19122}
\affiliation{Department of Biology, Temple University, Philadelphia, PA, 19122}

\author{Mohammad Erfan Mowlaei}
\affiliation{Institute for Genomics and Evolutionary Medicine, Temple University, Philadelphia, PA, 19122}
\affiliation{Department of Computer and Information Sciences, Temple University, Philadelphia, PA, 19122}

\author{Venkata Surya Kumar Choutipalli}
\affiliation{Institute for Computational Molecular Science, Temple University, Philadelphia, PA, 19122}
\affiliation{Temple Materials Institute, Temple University, Philadelphia, PA, 19122}
\affiliation{Department of Chemistry, Temple University, Philadelphia, PA, 19122}

\author{Mark DelloStritto}
\affiliation{Institute for Computational Molecular Science, Temple University, Philadelphia, PA, 19122}
\affiliation{Temple Materials Institute, Temple University, Philadelphia, PA, 19122}
\affiliation{Department of Chemistry, Temple University, Philadelphia, PA, 19122}

\author{Xinghua Shi}
\affiliation{Institute for Genomics and Evolutionary Medicine, Temple University, Philadelphia, PA, 19122}
\affiliation{Department of Computer and Information Sciences, Temple University, Philadelphia, PA, 19122}

\author{Micheal L. Klein}
\affiliation{Institute for Computational Molecular Science, Temple University, Philadelphia, PA, 19122}
\affiliation{Temple Materials Institute, Temple University, Philadelphia, PA, 19122}
\affiliation{Department of Chemistry, Temple University, Philadelphia, PA, 19122}

\author{Vincenzo Carnevale$^*$}
\affiliation{Institute for Computational Molecular Science, Temple University, Philadelphia, PA, 19122}
\affiliation{Institute for Genomics and Evolutionary Medicine, Temple University, Philadelphia, PA, 19122}
\affiliation{Temple Materials Institute, Temple University, Philadelphia, PA, 19122}
\affiliation{Department of Biology, Temple University, Philadelphia, PA, 19122}

\begin{abstract}$\null$\\$\null$\\{\bf ABSTRACT}\\
We present a variational autoencoder framework for learning and generating configurations of structured polymer globules from distance matrices. We used coarse-grained molecular dynamics to sample polyethylene structures, which we used as the training set for our deep learning model. 
By combining convolution and attention layers, the model encodes the structural patterns of distance matrices into an organized and roto-translationally invariant latent space of lower dimensionality. 
The generative capability of the variational autoencoder, coupled with a post-processing pipeline based on multidimensional scaling and short molecular dynamics, enables the recovery of physically meaningful configurations. The reconstructed and generated samples reproduce key observables, including energy, size, and entanglement, despite minor discrepancies in the raw decoder output. 
\end{abstract}
\date{\today}
\maketitle

\section{Introduction}

Deep-generative models are powerful tools for solving complex problems, and recently enabled breakthroughs in structural biology, natural language processing, material science, and other scientific and technological domains~\cite{DeepGenBioReview,choudhary2022recent,fuhr_deep_2022,deldjoo2024review,electronics13020322,du2024machine}. These models exploit the flexibility and expressive power of deep neural networks to fit high-dimensional probability distributions from a limited, yet extensive, number of examples and to generate new samples that preserve the structure and properties of the observed data~\cite{GenModRev1,GenModRev2}. 

In the field of material science, deep generative models have been extensively used to fit interaction potentials for atomistic simulations of solid-state and organic systems~\cite{lu_86_2021,qamar_atomic_2023,kovacs2025mace} and have been leveraged to bridge microscopic structures of amorphous systems with their macroscopic and thermodynamic properties~\cite{miccio2021mapping,kuenneth2021copolymer,martin2023emerging,yang2024novo,GlassTrans2025}. Among these amorphous materials, polymer melts represent a broad class of systems with relevance and potential for technological applications because of their desirable properties, such as flexibility, light weight, and strength~\cite{bracco2017ultra,chabi2019structure,ZHAO2020}. In these systems, the macroscopic properties emerge from the complex interplay between microscopic atomistic interactions and mesoscopic constraints imposed by chain connectivity, which often gives rise to patterns where regularity and randomness interact non-trivially and the structural organization encodes critical information about stress response, relaxation dynamics, and phase behavior~\cite{deGennesRep,klein1978evidence,TZOUMANEKAS200661,PrimPathRalf,tzoumanekas2006topological,PhysRevMaterials.3.015602}.

The structural characterization of these materials is typically pursued through standard techniques such as Molecular Dynamics and Monte Carlo simulations~\cite{KG1990,KalathiRouse2014,yamamoto_molecular_2013,hall2019polymer,binder1995monte,HsuMCMelt2014}, but both approaches face severe limitations in the high-density regime typical of melts, namely prohibitively long relaxation times in the former and low acceptance rates due to steric constraints in the latter, requiring the development of new strategies better suited to efficiently generate representative polymeric structures.

In the context of learning three-dimensional structural information, generative models have been highly successful in capturing the physical configurations of biomolecules. Many of these models are specifically designed to identify motifs in physical systems, and they have significantly expanded the repertoire of functional biological structures and small chemical compounds that can now be computationally generated without the need for lengthy design pipelines~\cite{abramson2024accurate,ESM3,ingraham2023illuminating,Ahern2025}. Extending their application to the generation of dense polymeric structures thus represents a promising avenue to advance our understanding and control of material properties, with potential applications in material design and defect detection~\cite{nilsson_modelling_2012,moyassari_molecular_2019,AMERI2024,BaiDef2025}.

Here, we explore this approach by training a Variational Autoencoder (VAE) on distance matrices computed from coarse-grained molecular dynamics simulations of a polyethylene globule at different temperatures. By designing the model as a combination of convolutional and attention layers, we leverage the roto-translational invariance and long-range character of the distance matrices to encode the system’s physical patterns into a low-dimensional latent space. Combining the generative capability of the VAE with a post-processing embedding algorithm for the distances, we demonstrate that the model can capture and generate physically meaningful configurations, which we then filtered and evaluated using key observables such as energy, size, and entanglement. Finally, we show that this generation procedure is computationally competitive with direct molecular dynamics relaxation, suggesting that canonical samples of dense polymeric systems can be efficiently embedded in a deep latent representation and subsequently generated on demand.

\section{Methods}

\begin{figure*}[t]
        \centering
        \includegraphics[width=6.6in]{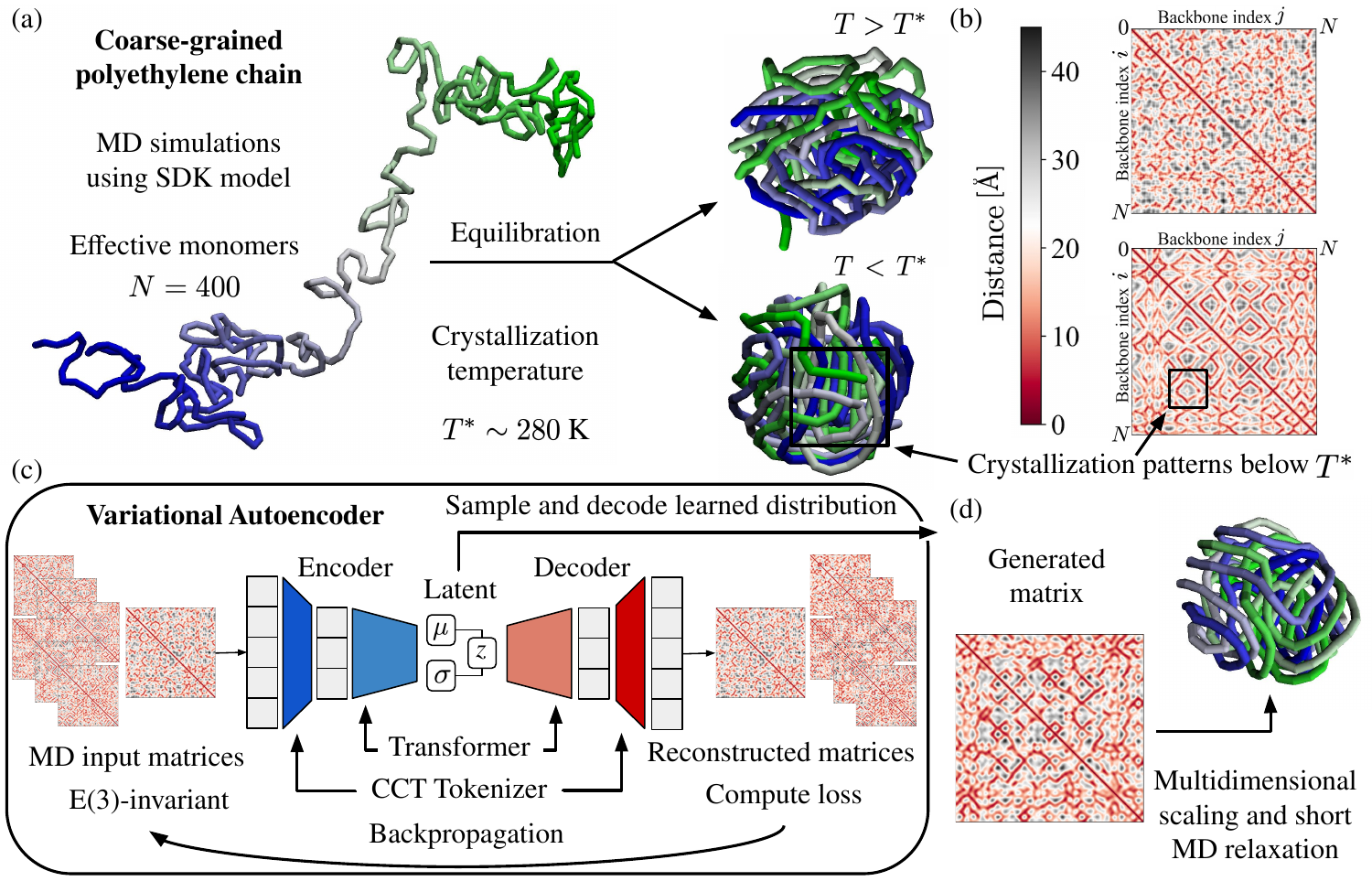}
        \caption{Molecular dynamics simulation setup and Variational autoencoder architecture. (a) Coarse-grained polyethylene chain with 400 effective monomers simulated via molecular dynamics using the SDK model. The system is initialized in a swollen state (left) and, after equilibration, reaches globular configurations (right) in both melted ($T$ $>$ $T$\(^{*}\)) and semi-crystalline states ($T$ $<$ $T$\(^{*}\)), with $T$\(^{*}\) indicating the nominal crystallization temperature of $\sim 280 \,\text{K}$. Monomers are colored according to their backbone index. (b) Distance matrices computed for the samples shown in panel a. Below $T$\(^{*}\), characteristic crystallization patterns arising from aligned PE stems appear. (c) Variational Autoencoder framework. Input distance matrices are first fed into an encoder composed of a CCT-based downsampling module and transformer blocks, which map them into latent variables ($\mu$, $\sigma$). Latent vectors $z$ are then sampled and passed through a decoder mirroring the encoder module to reconstruct distance matrices, which are finally used to compute loss functions and update the network weights. (d) Example of generated distance matrix obtained by sampling and decoding the learned latent distribution. Generated matrices are post-processed into 3D coordinates using multidimensional scaling and relaxed with short SDK simulations.}
        \label{fig:model}
\end{figure*}

\subsection*{Molecular Dynamics and Training Set}

We used the Shinoda-DeVane-Klein (SDK) model for polyethylene (PE) as a test of a generic polymer model~\cite{Shinoda}. Polyethylene is one of the simplest possible polymers, consisting of repeating methylene groups, and is relatively inert with no significant polar groups present. However, PE can still exhibit complex behavior due to its flexibility and semi-crystalline nature, and it has been extensively investigated using molecular dynamics simulations~\cite{RutledgeNucl2013,RutledgeLayer2017}. Nevertheless, significant controversy remains regarding the nucleation process of PE and the extent to which the structure of the nucleus influences crystallization~\cite{hall_monodisperse_2020}. These simulations typically require tens of microseconds of simulation time, thereby precluding the use of atomistic models.  We therefore use the coarse-grained SDK model where the polymer is represented as a series of beads connected by harmonic bonds and angles, such that each particle corresponds to three methylene groups.  Despite its simplicity, this model has been shown to reproduce the structural and thermodynamic properties of PE across a wide range of temperatures~\cite{hall_coarse-grain_2019}.

We generated configurations of a single chain in both melted and semi-crystalline states using molecular dynamics simulations.  In order to initialize each simulation, we created a self-avoiding random walk on a unit lattice and then scaled the system to the SDK bond length.  Although the random walk features right angles which are not commensurate with the SDK angle potential, this approach still produces a random distribution of initial conditions that, once relaxed, represent an entangled polymer (see Fig.~\ref{fig:model}a).  All subsequent simulations of the polymers were carried out with LAMMPS~\cite{plimpton_fast_1995,thompson_lammps_2022} using ``real'' units and a $5~{\rm fs}$ timestep.  Note that the true timescale of a coarse-grained system must be scaled due to the faster dynamics of the averaged potential, and in this case all times are given using a scaling factor of 4.07~\cite{hall_coarse-grain_2019}.  We then equilibrated the system by running the polymer at a 500 K for 10\(^{8}\) timesteps (\(\sim\)2 \(\mu\)s) before cooling the system down to the target temperature over the course of 10\(^{8}\) timesteps (\(\sim\)2 \(\mu\)s), yielding the globular states shown in Figure \ref{fig:model}a.  These conditions have been shown to yield crystallization in full melts of PE using the SDK model~\cite{hall_coarse-grain_2019}, and thus should yield crystalline domains within our single polymer.  Each polymer was then further equilibrated at the target temperature for 10\(^{8}\) timepsteps (\(\sim\)2 \(\mu\)s).  We chose three target temperatures in order to sample a range of crystallinities, namely 100, 200 and 300 K.

In order to identify the onset of crystallinity and provide data for training, we computed distance matrices for the polymer. For computational convenience, we simulated each polymer in a periodic box of length 20 nm. Since there are no long-range forces present in the model, this is large enough that the polymer chain is effectively isolated.  We thus computed distance matrices every 50000 timesteps by computing the absolute distance between each bead for the chosen frame.  These distance matrices yielded distinct patterns indicating aligned PE stems with ordered loops in the globule for \(T<T^*\), as shown in Fig.~\ref{fig:model}a,b and Fig.~S1. 

\subsection*{Machine Learning Model}

{\bf Architecture and Loss Function.} Our machine learning model leverages the roto-translational symmetry of distances and the physically nonlocal information encoded in distance matrices to reduce the learning of structures to the identification of patterns within the matrices, which are treated as images where each pixel $(i,j)$ corresponds to the distance between atoms $i$ and $j$ in the polymer. Sketches of the architecture are shown in Fig.~\ref{fig:model}c and Fig.~S2 and the code is available on the GitHub repository \cite{githubPolymer_VAE}.

Following the approach of the Compact Convolutional Transformer \cite{HassaniWaltonCCT2022,mowlaei2025stici}, we designed the encoder of our VAE \cite{VAE} to first transform the input matrix, $x$, into a sequence of tokens using three 2D convolutional layers with kernel sizes of $7$. Each layer maps a subregion of the matrix, together with its temperature auxiliary input, into a token vector that captures the nonlocal patterns in the corresponding region of the distance matrix. The subregion is shifted along the matrix with a stride of $2$ in both dimensions, resulting in three successive dimensionality reductions that map the original matrix of shape $N\times N$ into a tokenized vector of shape $256 \times N/(2*2)^3$, where $256$ represents the custom number of output features. This vector is then passed through six one-dimensional convolutional layers with varying kernel sizes $(3,5,7,7,15,3)$, followed by a multi-head attention transformer with $32$ attention heads \cite{Transformers}. This procedure does not change the dimensionality of the vector but allows the model to incorporate information from larger scales of the matrix and to capture relevant features in the tokens.
Finally, the mean and logarithmic variance of the latent representation are computed via two fully-connected layers that map the output of the transformer into two latent vectors, $\mu$ and $\log \sigma^2$, each of dimension $1200$. The components of these vectors after training are shown in Fig.~S3. This dimensionality was validated \emph{a posteriori} based on the spectra of the latent representation, which show that the eigenvalues of the latent PCA converge to zero when the dimension approaches $1200$ (see Fig.~S4).

After this encoding step, a point $z = \mu + \sigma \odot \epsilon$ is sampled in the latent space, where $\epsilon \sim \mathcal{N}(0, I)$ is a mulivariate standard normal and $\odot$ denotes element-wise multiplication. This latent vector is stacked with the corresponding input temperature and passed to the decoder, whose architecture mirrors that of the encoder, i.e., it consists of a sequence of transformer block, 1D convolutions, and 2D convolution layers with upsampling. The output of the decoder is finally passed through a transpose convolution layer, which reshapes the matrix to the correct form $N \times N$ and enforces zero diagonal, positivity of the entries, and symmetry of the reconstructed point $\hat{x}$.

Following the standard approach of \cite{VAE}, the input $x$ and its decoded image $\hat{x}$ are used to compute the evidence lower bound
\begin{equation}
\mathcal{L}_{\text{ELBO}} = \mathbb{E}_{q(z|x)}[\log p(x|z, T)] - \beta \cdot \mathrm{KL}(q(z|x) \, || \, p(z)),
\end{equation}
\noindent where the first term is the reconstruction loss, which drives $\hat{x}$ towards the input $x$, while the second term is the Kullback–Leibler divergence between the approximate posterior $q(z|x)$ and the multivariate Gaussian prior $p(z) = \mathcal{N}(0, I)$, which favors continuous coverage of the latent space. The weighting coefficient $\beta = 0.1$ was set to balance the relative contributions of the two terms during training, as first proposed in \cite{betaVAE}. 

The KL term can be computed analytically using $\mu$ and $\sigma$ due to the Gaussian form of both the prior and the approximate posterior~\cite{VAE}, while the reconstruction term was modified to emphasize relative differences in the reconstructed matrix $\hat{x}$. Specifically, instead of the standard mean-squared error (MSE) between $x$ and $\hat{x}$, we defined
\begin{equation}
    \log p(x|z, T) := {\rm MSE}\!\left(\frac{\hat{x}(z,T)}{x(T)}, \mathbf{1}\right),
\end{equation}
\noindent where $\mathbf{1}$ is the matrix of all ones and the division is intended element-wise, excluding the zero diagonals, whose ratio is set to 1. In this way, errors on small input distances are penalized more strongly, and the model focuses on short-range patterns in the matrices, which are more sensitive to the physical structure of the globule. The average over $q(z|x)$ in the reconstruction term is estimated using a mini-batch average with batch size 8, and the batch-averaged loss is used to update the model's weights via backpropagation. 

The model was trained on a single NVIDIA 80GB A100 GPU for 110 epochs using the Adam optimizer with an initial learning rate of $4 \times 10^{-7}$, and a learning rate scheduler with patience of $3$ epochs and minimum learning rate of $1 \times 10^{-12}$. Early stopping based on the validation reconstruction loss was applied to prevent overfitting, with a patience of $10$ epochs. The progression of the losses is reported in Fig.~S5.

{\bf Sample Generation and Postprocessing.} After training is completed and the network weights have converged to their optimal values, the encoder can be used to obtain the final latent representation of the training set, which defines the distribution of points in the latent space. As shown in Fig.~S6–S7, these points occupy a limited region of the 1200-dimensional space, as a consequence of the KL regularization term in the loss function, and are stratified along the first principal component according to the temperature associated with each matrix, suggesting that the model successfully incorporates temperature information to distinguish between the different structural patterns encoded in the matrices.

In addition, given a point in the latent space and an associated temperature $T \in \{100, 200, 300\}$ K, whether it is originating from the training set or sampled directly from the prior distribution $p(z)$, the decoder can reconstruct a distance matrix that, in principle, corresponds to a polymer configuration. Since the decoder is trained on the encoded training points, the realism of a decoded matrix is expected to increase with the proximity of the latent point to those originating from the training data.

To improve the quality of the generated samples, we did not draw new points in the latent space from a simple Gaussian prior, but rather from a Gaussian mixture model (GMM) with 300 full-covariance components, fitted \emph{a posteriori} to the latent distribution of the training set.

Finally, as shown in Fig.~\ref{fig:model}d, once a newly sampled latent point is decoded, the generated matrix is passed through an embedding algorithm to recover the final configuration of the system. Specifically, the matrix is processed using classical Multidimensional Scaling (MDS), which diagonalizes the Gram matrix associated with the distance matrix and computes its square root to obtain the coordinates of $N$ three-dimensional points. Since the generated matrix is not guaranteed to satisfy all the properties of a distance matrix, as discussed in the following paragraph and in the Results, the reconstructed coordinates may display non-physical features such as atomic overlaps and sharp bends in the backbone. To mitigate these artifacts, we subsequently performed an energy minimization followed by a short MD simulation using the SDK potential (see Molecular Dynamics and Training Set), thereby ensuring that the resulting configurations can be legitimately compared with the physical inputs.

\subsection*{Observables} 
We used the following mathematical and physical observables to assess the quality of the matrices generated by the model.  

{\bf Triangle inequality.} From a mathematical perspective, the outputs of the decoder should represent distance matrices, however, not all properties of a metric are necessarily satisfied. While positivity and symmetry were enforced by the final layer of the decoder (see the section Machine Learning Model), the triangle inequality was not explicitly imposed due to the high computational cost of verifying it: the inequality must hold for all triplets of distances $d_{ij}, d_{jk}, d_{ik}$, which scales as $N^3$. Consequently, the model must learn by itself to satisfy the triangle inequality during training. To quantify this ability, we evaluate the degree of violation using
\begin{equation}
    v = \frac{d_{ij}+d_{jk}}{d_{ik}},
    \label{eq:triangle}
\end{equation}
\noindent which must be greater than 1 for every valid triplet of distinct points $i,j,k$.

{\bf Radial distribution function.} We used the radial distribution function $g(r)$ of the globule to assess how well the model captures the structural features of the system. Since $g(r)$ depends only on pairwise distances, it can be computed directly from the decoder output without requiring the additional embedding step in physical space described in the Methods. This is done by rescaling the probability density of the distances with the distance itself $g(r) := P(r=d)/d$. Note that, due to the finite system size, all $g(r)$ are expected to converge to zero for $r$ larger than the system diameter.

{\bf Potential energy.} We leveraged the MD simulations used to construct the training set to obtain direct access to the potential energy of the configurations. After embedding, the coordinates were used to evaluate the SDK potential (see section Molecular Dynamics and Training Set) on the decoder output, and the resulting values were used as an assessment criterion for the quality of the model. It is important to note that, due to the presence of the angular term in the potential, the embedding constitutes a necessary intermediate step for the energy calculation.

{\bf Writhe.} Since the system is an open chain, its topological state is not strictly defined; however, its degree of entanglement can be quantified through the writhe. To compute it, we employed a stochastic algorithm in which the configuration is projected onto many (100) random planes, and the mean number of signed crossings is evaluated for each projection. The sign of a crossing is determined by the orientation of traversal along the curve~\cite{van1998writhe}. The average of these mean signed crossings over all projections provides a quantitative measure of the extent to which the globule twists around itself. As in the case of the potential energy, the embedding constitutes a necessary preliminary step for the computation of this observable.

{\bf Radius of gyration.} As is customary in polymer physics, we used the radius of gyration $R_g$ to quantify the size of the globule. Although $R_g$ can also be computed directly from the distance matrix, we evaluated it from the embedded coordinates of the decoder output in order to place this observable on the same footing as the others measured physical observables.

\section{Results}

\subsection*{Reconstructed samples}
We first evaluate the ability of the model to reconstruct input configurations after training. To do this, we provided the trained VAE with an input matrix $x$ from the training set and computed its encoder-decoder image $\hat{x}$. The two matrices, $x$ and $\hat{x}$, can then be directly compared through their physical observables, or alternatively, ensemble averages over many input-reconstruction pairs can be examined. Examples of reconstructed samples corresponding to the inputs shown in Fig.~S1 are reported in Fig.~S8.

\begin{figure}[t]
        \centering
        \includegraphics[width=3.3in]{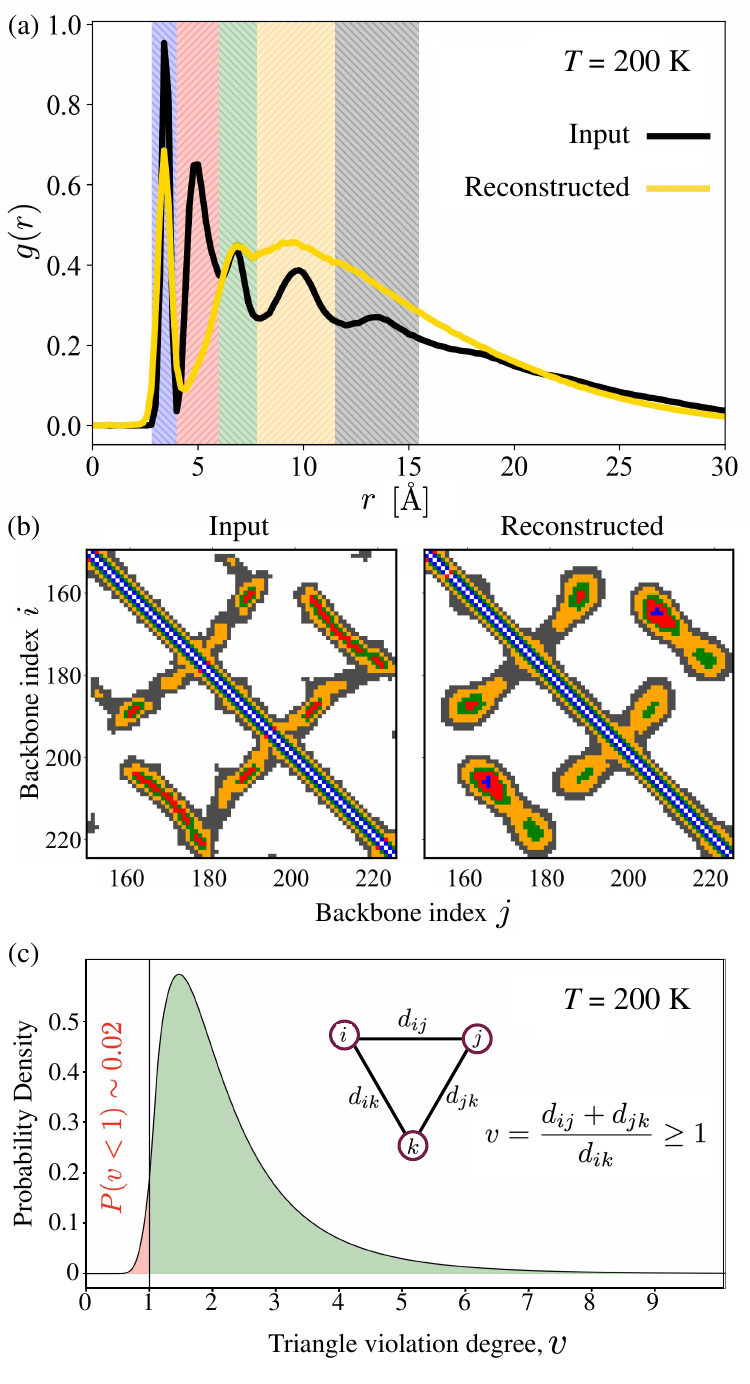}
        \caption{Reconstruction of structural patterns from input distance matrices. (a) Radial distribution functions $g(r)$ of input (black) and reconstructed (yellow) matrices at $T = 200$ K. The main peaks of the distribution are highlighted by shaded regions. (b) Example submatrices ($70 \times 70$) extracted along the main diagonal of an input (left) and the corresponding reconstructed (right) distance matrix, with elements colored according to the $g(r)$ peaks in panel (a). (c) Probability distribution of the triangle inequality violation degree $v$ for all reconstructed samples at $T = 200$ K.}
        \label{fig:rec1}
\end{figure}

\begin{figure*}[t]
        \centering
        \includegraphics[width=6.6in]{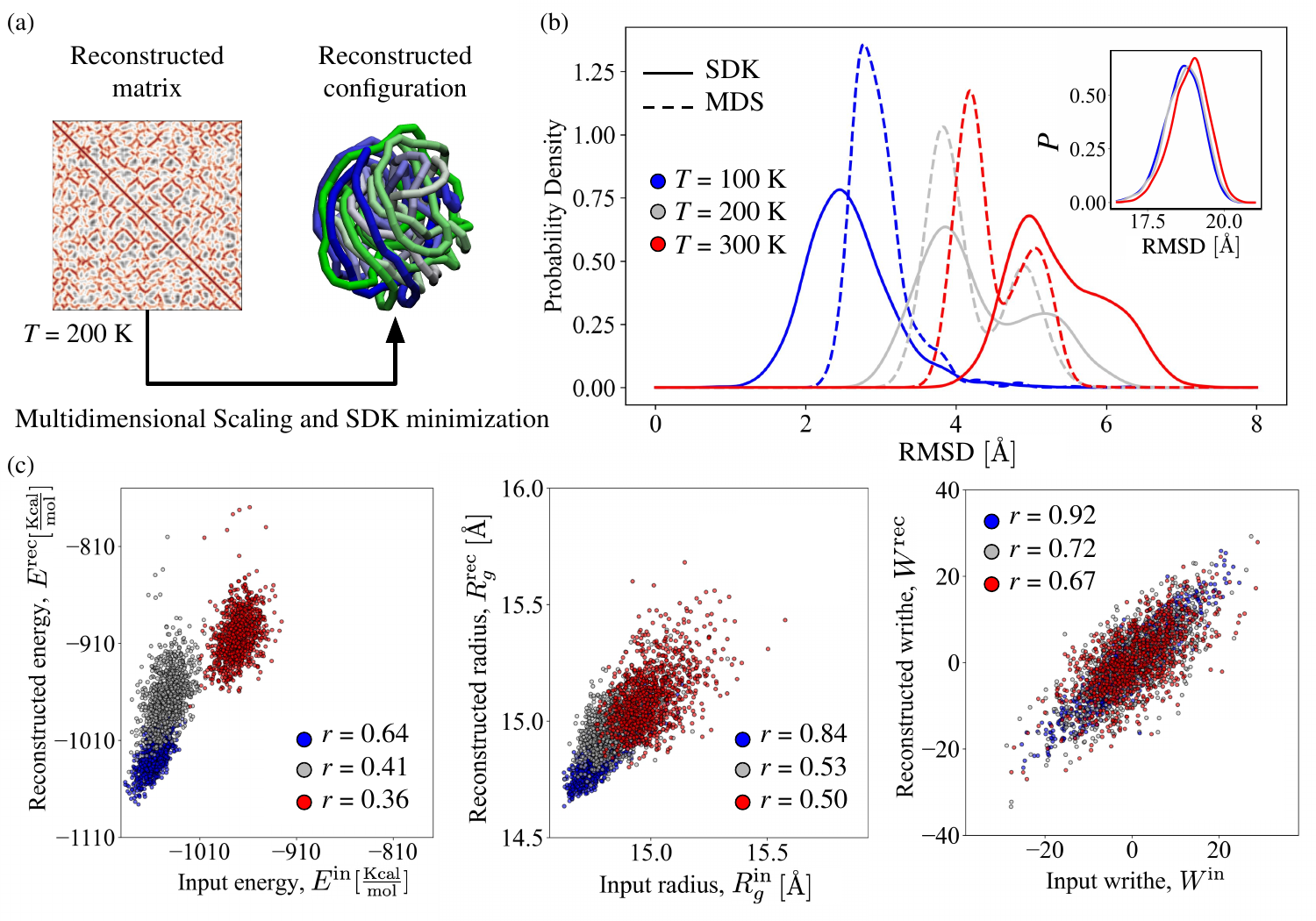}
        \caption{Properties of embedded reconstructed samples. (a) Example of reconstructed distance matrix at $T = 200$ K, post-processed via multidimensional scaling (MDS) and SDK energy minimization, yielding a three-dimensional configuration corresponding to the nearest local energy minimum. (b) Probability distributions of the root-mean-squared distance (RMSD) between optimally aligned reconstructed and input configurations for the three explored temperatures, with SDK-minimized configurations (solid lines) compared to raw MDS embeddings (dashed lines). In the inset, same distributions computed using uncorrelated canonical samples from MD simulations. All the distributions are smoothed using a Gaussian KDE for visual clarity.  (c) Scatter plots comparing input and reconstructed physical observables: potential energy $E$ (left), radius of gyration $R_{g}$ (middle), and writhe $W$ (right). Points are colored by input temperature (same scheme as panel b), with Pearson correlation coefficients $r$ reported for each temperature.}
        \label{fig:rec2}
\end{figure*}

Fig.~\ref{fig:rec1}a shows the radial distribution functions $g(r)$ of the input and reconstructed samples at $T = 200~\text{K}$, computed directly from the distance matrices as described in the Methods section. The input distribution (black line) exhibits a complex structure with multiple peaks (highlighted by the shaded regions in the figure), whereas the reconstructed distribution (yellow line) appears smoother and less structured. In particular, the model accurately reproduces the first and third peaks of the input distribution, centered at $3 \text{\AA}$ and $6.5 \text{\AA}$, while the other features are partially suppressed during the encoding–decoding process. Notably, the pronounced peak around $5 \text{\AA}$ in the input distribution is absent in the reconstructed case, and distances in the range $7 \text{\AA}$–$20 \text{\AA}$ are systematically overrepresented. At larger distances, beyond $20 \text{\AA}$, comparable to the diameter of the system as discussed below, the two distributions are in close agreement.

To gain further insight into the origin of the different peaks, Fig.~\ref{fig:rec1}b presents an example of an input–reconstruction pair of submatrices of size $70 \times 70$ extracted along the main diagonal. Each cell is colored according to the shaded regions associated with the peaks in the input $g(r)$ shown in panel a. 

The results indicate that the first peak (blue) arises primarily from the first off-diagonal elements of the matrix, which correspond to next-nearest distances along the polymer backbone and are in agrrement with the nominal bond length of the SDK potential~\cite{hall_coarse-grain_2019}. This feature is consistently reproduced in the reconstructed matrix.
In contrast, the second peak (red), which is entirely suppressed by the reconstruction procedure, is related to ordered structures of the globule reminiscent of semicrystalline patterns in polyethylene. In the input matrix, this contribution originates from single-pixel-wide lines running parallel or orthogonal to the main diagonal. These patterns suggest that the corresponding distances arise from backbone segments positioned close to each other within the globule. The connection between the red peak and semicrystalline ordering is further supported by the sharpening of this peak at lower temperatures, as shown in Fig.~S9. Unfortunately, these patterns are not faithfully reproduced in the reconstructed samples: they are underrepresented and lose the regular character observed in the input, appearing instead as small, patch-like regions of nearly constant distance, exemplified by the red elliptical features in the reconstructed panel (right). However, as we demonstrate below, the embedding procedure mitigates this limitation, and the final configurations recover all the relevant physical patterns.
The third (green) and fourth (yellow) peaks, which are only partially captured by the model, are associated with secondary ordering structures that emerge from the red patterns, as evidenced by the way the corresponding regions in the input matrix are layered around the crystalline lines. Finally, the last peak (black) together with the white background accounts for all pairwise distances exceeding the characteristic structural scales of the globule, with $g(r)$ going to zero as $r$ approaches the system size.

As a final evaluation of the quality of the decoder reconstruction, we computed the degree to which the reconstructed matrices satisfy the triangle inequality. While enforcing positivity and symmetry in the model is straightforward, explicitly imposing the triangle inequality within the loss function is computationally prohibitive and therefore was not implemented. Nonetheless, the decoder output satisfies this constraint with reasonable accuracy, as indicated by the distribution of the violation parameter $v$ (see Methods, Eq.~\ref{eq:triangle}) in Fig.~\ref{fig:rec1}c. For true distances, $v$ must exceed 1, and we find that the reconstructed matrices violate this condition in only about $2\%$ of cases, a behavior that remains consistent across temperatures (Fig.~S10).

Overall, while the decoder output shows some loss of fine structure in the radial distribution function and minor violations of the triangle inequality, the model reproduces the main features of the matrix. Moreover, as we demonstrate later, the embedding procedure recovers most of the relevant physical properties of the system.

\begin{figure*}[t]
        \centering
        \includegraphics[width=6.6in]{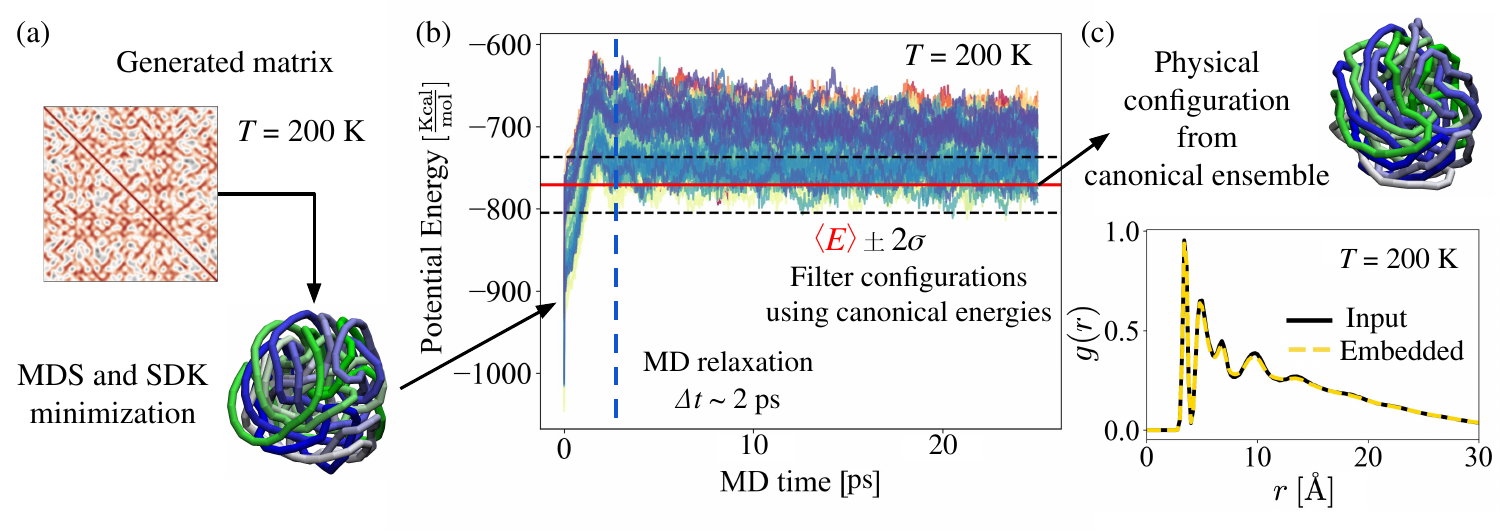}
        \caption{Generation and embedding of novel samples. (a) Example of generated distance matrix at $T = 200$ K embedded by multidimensional scaling and minimized with the SDK potential. (b) Potential energy trajectories during MD relaxation of $100$ generated samples at $T = 200$ K. Convergence is reached on the order of 3 ps (vertical dashed line), but the stationary energy distribution is broader than that of the canonical ensemble. The ensemble mean $\langle E \rangle$ (solid red line) and standard deviation $\sigma$ (dashed black line) are indicated, and only configurations with relaxed energies within $\langle E \rangle \pm 2\sigma$ are kept. (c) Final generated configurations after the filtering procedure described in panel b. These configurations reproduce the radial distribution function $g(r)$ in agreement with the corresponding input at $T = 200$ K.}
        \label{fig:gen1}
\end{figure*}

After evaluating the quality of the reconstructed matrices at the decoder output, we applied our embedding procedure (detailed in the Methods section) to obtain three-dimensional physical configurations that can be directly compared with the input configurations from the MD simulations. This procedure combines classical multidimensional scaling with subsequent energy minimization of the generated coordinates, yielding configurations corresponding to the nearest local energy minimum associated with the reconstructed matrix, as illustrated in Fig.~\ref{fig:rec2}a. For a fair comparison, we also minimized the energy of the input configurations, which were originally obtained from finite-temperature MD simulations, and aligned them to the reconstructed configurations using the Kabsch algorithm. It is important to note that distance matrices do not contain information about the parity of the system. Therefore, for each reconstructed matrix we performed the analysis using both the embedded configuration and its mirror image, retaining the case that provided the best agreement. 

The probability density of the root-mean-squared distance (RMSD) between the optimally aligned reconstructed and input configurations is shown in Fig.~\ref{fig:rec2}b. Across all temperatures, the RMSD values remain below $8 \text{\AA}$, which is approximately one third of the RMSD measured between independent samples drawn from the canonical ensemble (inset of Fig.~\ref{fig:rec2}b). Furthermore, configurations obtained after SDK minimization (solid lines in the figure) at $T=100$ K show improved accuracy compared to those reconstructed using MDS alone (dashed lines). At higher temperatures, however, the effect is less consistent, with mixed results at $T=200$ K and a clear deterioration at $T=300$ K, where a pronounced bump at larger RMSD values emerges.
The center of the RMSD distributions also depends on the simulation temperature, with larger RMSD values observed at higher temperatures. This can be rationalized by noting that at lower temperatures, semi-crystalline motifs emerge in the distance matrices, providing the VAE with richer structural information for building its latent representation. 

It should be emphasized that the simulation temperature influences only the reconstruction accuracy of the model and does not explicitly appear in the final minimized configurations used for comparison. Indeed, for genuine physical configurations, such as the canonical ensemble benchmarks shown in the inset, all values of $T$ yield the same RMSD distribution after minimization.

Finally, we compared the input and reconstructed values of three physical observables, namely the potential energy $E$, the radius of gyration $R_g$, and the writhe $W$ (see Methods). These comparisons are reported in the scatter plots of Fig.~\ref{fig:rec2}c, where points are colored according to their input temperature, with the associated marginal probability distributions shown in Fig.~S11.
The reconstructed energies, while spanning ranges comparable to those of the inputs, are not well correlated with them (with linear correlation coefficients, $r$, in the range $0.36 < r <0.64$) and are systematically higher across all explored temperatures. This discrepancy can be understood in light of the multidimensional scaling procedure used to transform a distance matrix into a set of coordinates. Since the reconstructed matrices contain no explicit energetic information and are only approximations of real distance matrices (see Fig.~\ref{fig:rec1}c), the resulting MDS configurations often display overlapping atoms and backbones that are either unnaturally straight or exhibit sharp bends. Although the subsequent energy minimization step alleviates these artifacts, the introduction of steric interactions can trap the system in conformations where bond-bond angles are strongly constrained, leading to minimum-energy states that differ from those of the corresponding input configurations. This effect explains the systematic discrepancy observed between reconstructed and input potential energies.
This feature is less pronounced in metric and topological observables such as the radius of gyration $R_g$ and the writhe $W$, which display good correlations with their input values. In particular, the accuracy of $R_g$ improves with decreasing temperature, reaching $r \sim 0.84$ for $T=100$ K, while $W$ is well reproduced ($0.67 < r <0.92$) and remains nearly independent of $T$.

\subsection*{Generated samples} 

We next analyzed the ability of the VAE to generate new physical configurations by sampling the latent space and subsequently decoding the sampled points, following the procedure illustrated in Fig.~\ref{fig:gen1}. Specifically, we first sampled the latent space using a Gaussian mixture model (see Methods) and passed the sampled points to the decoder, which produced corresponding matrices. These matrices were then embedded into three-dimensional coordinates using MDS, followed by SDK minimization, as done for the reconstructed samples. To account for the fact that the minimized configurations do not explicitly contain temperature information, present only in the latent space representation (see Fig.~S6) but lost during energy minimization, we further refined the structures through short MD simulations. 

Figure~\ref{fig:gen1}b shows the potential energy trajectories during MD relaxation for 100 generated matrices at $T=200~\mathrm{K}$, each initialized from the configurations obtained after the MDS and SDK minimization steps. The results indicate that the typical relaxation time required for the generated globules to reach convergence is on the order of $\Delta t \sim 2~\mathrm{ps}$. However, the stationary energy distribution of the generated samples exhibits a broader spread than the corresponding physical distribution obtained from the canonical ensemble, whose mean $\langle E \rangle$ and variance $\sigma$ are indicated in the figure by the solid red and dashed black lines, respectively, while the full distributions are reported in Fig.~S12. To ensure physical consistency, we retained as final embedded configurations only those whose relaxed energies fell within the range $\langle E \rangle \pm 2\sigma$.

\begin{figure}[t]
        \centering
        \includegraphics[width=3.3in]{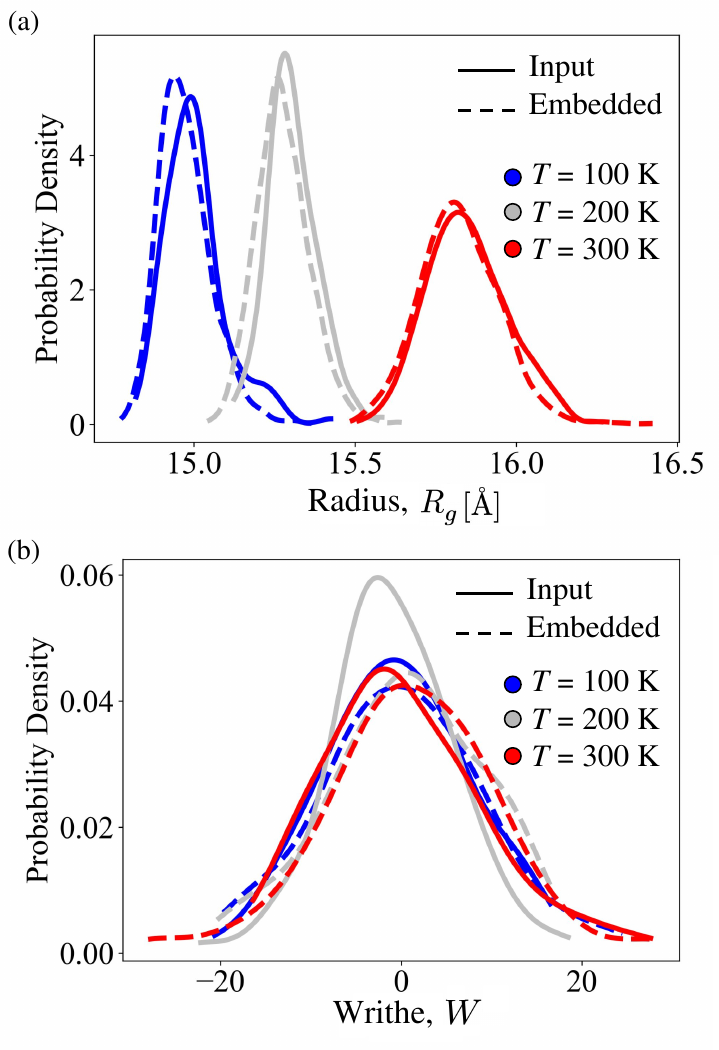}
        \caption{Physical observables of embedded generated samples. Probability distributions of (a) the radius of gyration $R_{g}$ and (b) the writhe $W$ for input (solid lines) and generated embedded (dashed lines) configurations at the three explored temperatures. All the distributions are smoothed using a Gaussian KDE for visual clarity and curves are colored by input temperature.}
        \label{fig:gen2}
\end{figure}

Figure~\ref{fig:gen1}c shows the final outputs of this procedure. These configurations have physical energies by construction, and they also reproduce the radial distribution function $g(r)$ in excellent agreement with the corresponding input temperature. The yield of acceptable configurations depends strongly on the temperature, ranging from only $13\%$ at $T=100$ and $200$ K, to $97\%$ at $T=300$ K (see also Fig.~S13 for energy trajectories and Fig.~S14-S15 for radial distributions before and after embedding at additional temperatures). Examples of both physical and nonphysical generated matrices, together with the corresponding distributions of triangle inequality violations, are reported in Figs.~S16–S19. 

Distance matrices recomputed from the embedded physical configurations are displayed in Fig.~S20, where semicrystalline patterns and their emergence at decreasing temperature are clearly visible. Unfortunately, we were not able to identify any feature that distinguishes the two groups \emph{a priori}, without performing the full embedding procedure.

We also measured the probability densities of the radius of gyration and the writhe for the embedded configurations (Fig.~\ref{fig:gen2}), finding a good overlap between the final outputs of our model and the canonical distributions computed from the training set at different temperatures. The accuracy remains stable across temperatures, with $R_g$ increasing with $T$, while the writhe $W$ remains nearly constant, consistent with the results obtained for the reconstructed samples.

From a computational standpoint, the MD time cost for relaxing a generated sample ($\sim 2~{\rm ps}$) is approximately one order of magnitude smaller than the cost of relaxing a swollen self-avoiding walk configuration ($15~{\rm ps}$). Unfortunately, the low generative efficiency at low temperature ($\sim 13\%$) makes the overall generation cost comparable to that of a full molecular dynamics approach. However, it is worth noting that the relaxation time of full MD is expected to scale exponentially with system size, whereas the time required to adjust a raw generated sample, already close to a representative configuration, is expected to scale more slowly. Notably, sampling from the latent GMM and decoding the outputs incurs a computational cost one order of magnitude smaller than the one of the MD step on our hardware (AMD EPYC 7543 32-Core CPU) and therefore has a negligible impact on the overall efficiency. 

\begin{figure}[b]
        \centering
        \includegraphics[width=3.3in]{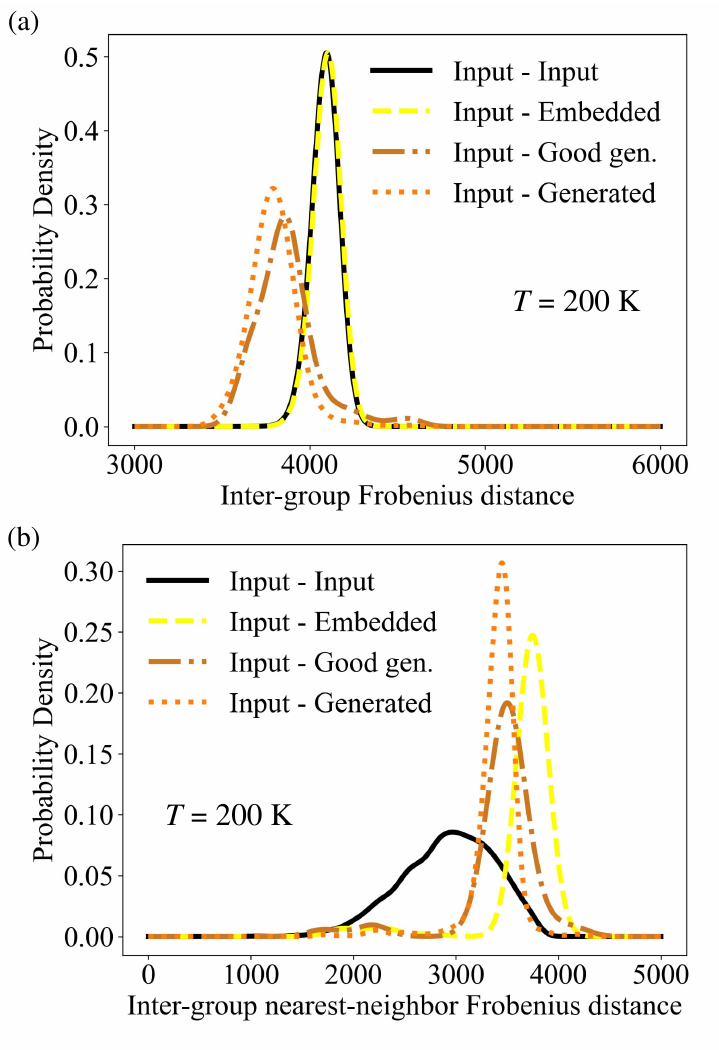}
        \caption{Novelty analysis of generated samples. (a) Probability distributions of Frobenius distances at $T = 200$ K between pairs of matrices from different groups: input-input (black), input-generated raw decoder outputs (orange), input-energy filtered outputs (brown), and input-embedded configurations (yellow). (b) Probability distributions of nearest-neighbor Frobenius distances at $T = 200$ K for the same groups as in panel a. All distributions in both panels are smoothed using a Gaussian KDE for visual clarity.}
        \label{fig:gen3}
\end{figure}

Finally, we evaluated the novelty of the generated samples to quantify the ability of latent-space sampling to generalize beyond the training set and produce configurations that are not only physically relevant but also distinct from those already observed. To this end, we first divided a set of input matrices into two random groups and computed the inter-group Frobenius (Euclidean) distance as a reference. We then generated a set of matrices and calculated their Frobenius distances with respect to one of the two input subgroups. The same analysis was repeated for generated matrices that will survive the energy filtering after the embedding procedure and, finally, for their embedded counterparts obtained after the MD relaxation.

Fig.~\ref{fig:gen3}a shows the distributions of the inter-group distances for $T= 200$ K, from which two key features emerge. First, the input–input pairs (black solid line) and the input–embedded pairs (yellow dashed line) are perfectly superimposed, indicating that the generated matrices are statistically indistinguishable from the inputs. This strongly supports their ability to correctly represent physical polymer configurations. Second, this superposition is recovered only after the embedding procedure: the raw decoder outputs (orange dotted line) and the decoder outputs that survive energy filtering (brown dash-dotted line) display clearly separated distance distributions relative to the training set. It is worth noting that these distributions cannot themselves be used as a selection criterion for viable generated samples, owing to their substantial overlap. 

Panel b of Fig.~\ref{fig:gen3} shows the same distance distributions, but computed using only the nearest-neighbor matrices. Interestingly, the input–embedded distribution is shifted to the right of the input–input distribution, indicating that GMM sampling effectively explores the matrix space and that the decoder produces samples beyond simple replicas of the training data. As before, the generated and “good generated” distributions overlap and therefore cannot serve as a selection criterion. These qualitative features are consistently observed across all investigated temperatures (Figs.~S21–S22). For completeness, the same analysis performed in the latent space prior to decoding is reported in the Supporting Information (Figs.~S23–S24), showing the same behavior as in the decoded (real) space and suggesting that the latent representation is approximately distance-preserving, up to a scaling factor.

Our results support the fact that the generation process is effectively able to provide novel configurations that are also statistically indistinguishable from canonical samples of the system.

\section{Discussion and Conclusions}
We designed a variational autoencoder architecture to learn dense polymer configurations from distance matrices, using molecular dynamics simulations of a coarse-grained polyethylene as the training set. Our approach leverages the E(3) invariance of distances and the non-local nature of structural patterns in distance matrices to construct a structured latent space that preserves the essential physical features of the system.

To evaluate the learned latent representation, we performed two levels of analysis. First, we assessed the ability of the trained model to reconstruct input matrices. Second, we examined the physical relevance of newly generated configurations obtained through probabilistic sampling of the latent space. In both cases, the model’s performance was not flawless: the raw decoder output exhibited minor violations of the triangle inequality as well as discrepancies in the monomer radial distribution function.

However, the retained information was sufficient to recover physically accurate configurations through a post-processing pipeline combining multidimensional scaling with short molecular dynamics simulations. This procedure yields viable configurations that reproduce the correct statistical properties of energy, size, and entanglement.

Our results demonstrate that the model can be used to encode molecular dynamics data in a low-dimensional latent space, which can subsequently be sampled to generate diverse and physically viable configurations. In the system under analysis, this approach presents a computational cost comparable to that of standard molecular dynamics simulations alone, with the potential of a better scalable strategy for studying polymer systems.

Different directions can be explored to improve the design and performance of the model. One promising direction is to modify our model to operate directly on coordinates. While distance matrices naturally possess roto-translational symmetries that simplify the construction of a latent representation without requiring the model to learn these symmetries during training, they also scale quadratically with the number of atoms, making memory intensive the extension of the model to larger systems. By employing equivariant versions of the model components~\cite{bronstein2021geometric}, it may be possible to encode these symmetries directly into the architecture of the VAE rather than relying on their presence in the training set.

Moreover, it will be important to design a model capable of handling systems in the dense phase with periodic boundary conditions, rather than only globular isolated systems. This extension would enable the application of our approach to more general scenarios and potentially make it suitable for practical contexts such as finite element simulations.

Finally, an interesting direction for future work is to investigate the applicability of our method to atomistic simulations, where the reduction in computational cost could significantly enhance the availability of fine-grained configurations of polymeric materials.

\section{Acknowledgements}
We are grateful for Invaluable high performance computing resources provided by the US National Science Foundation via Major Research Infrastructure grant numbers: 1625061 and 2216289. Research by VC, MDS, and MLK was sponsored by CCDC-ARL and was accomplished under Cooperative Agreement Number W911NF-21-2-0007. The views and conclusions contained in this document are those of the authors and should not be interpreted as representing the official policies, either expressed or implied, of the U.S. Army Research Laboratory or the U.S. Government.

\bibliography{bibliography}

\end{document}


\sffamily

\centerline{\Large Supporting Information}

\vskip 0.2cm
\centerline{\Large \shortstack{Generative Modeling of Entangled Polymers with a \\ Distance-Based Variational Autoencoder}}
\vskip 0.2cm
\centerline{\large \shortstack{by P.~Chiarantoni, O.~Serra, M.E.~Mowlaei, V.S.K.~Choutipalli, \\ M.~DelloStritto, X.M.~Shi, M.L.~Klein and V.~Carnevale}}
\vskip 1.5cm

\begin{figure}[h]
        \centering
        \includegraphics[width=7in]{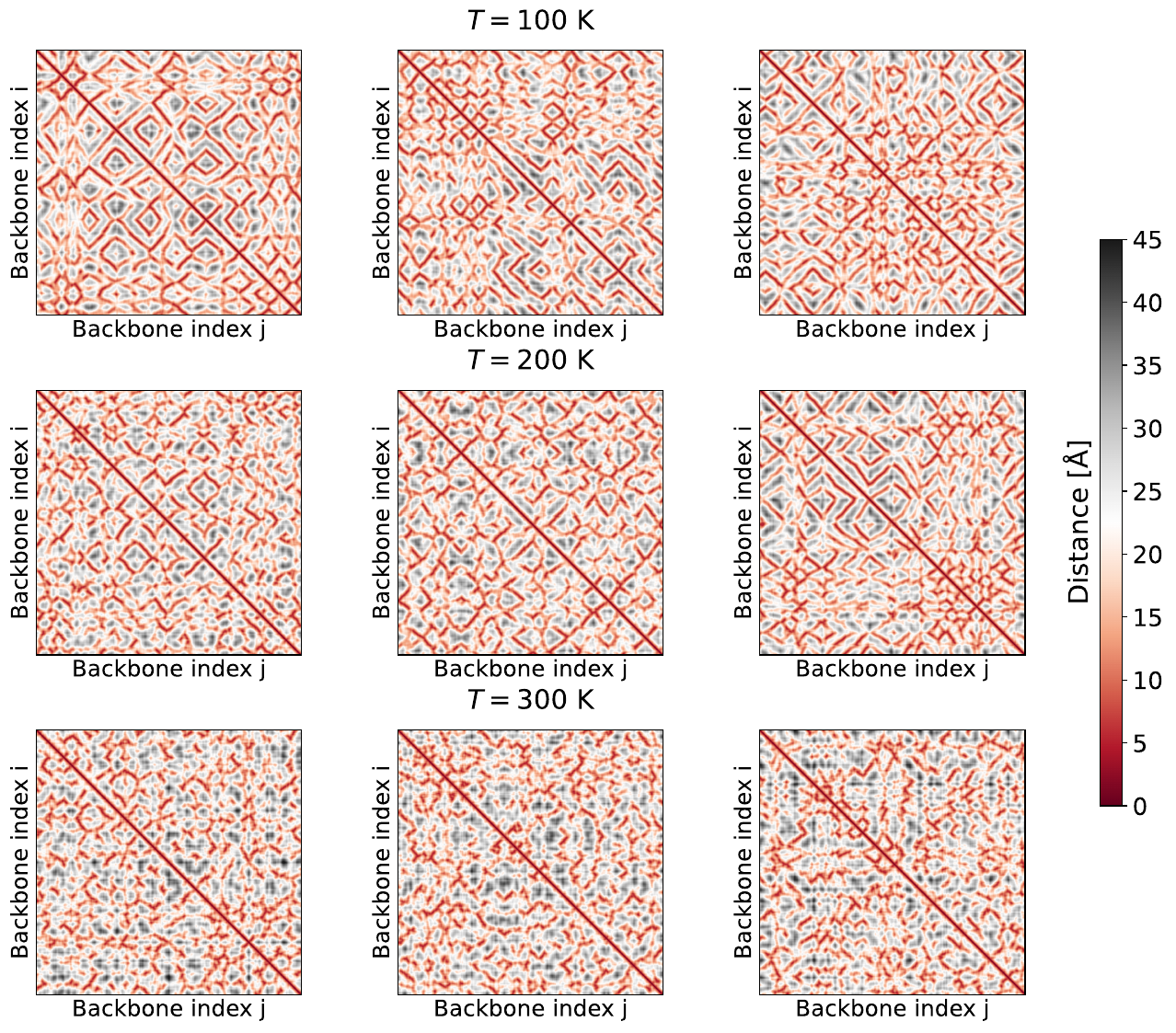}
        \caption{Examples of input distance matrices of polyethylene chains at $T = 100$, $200$, and $300$ K. It can observed how the ordered semicrystalline patterns progressively reduce as the temperature increases.}
        \label{fig:S1}
\end{figure}

\begin{figure}[t]
        \centering
        \includegraphics[width=7in]{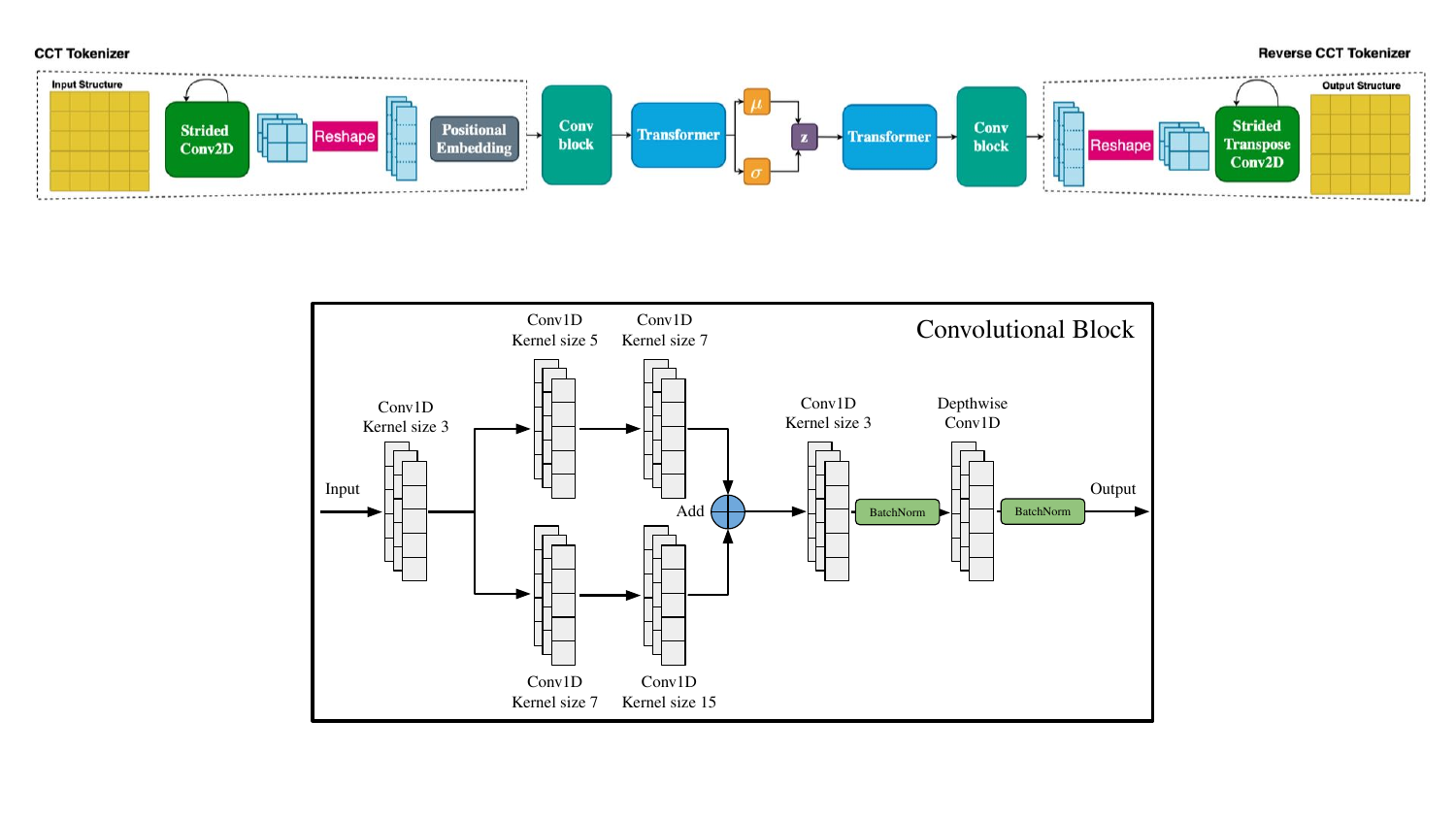}
        \caption{Architecture of the variational autoencoder (VAE) used in this work. Top panel shows the full encoder–decoder pipeline, including the CCT tokenizer, convolutional block, and Transformer layers, which map the input distance matrices into a latent distribution $(\mu, \sigma)$ and reconstruct them through a symmetric decoder with reverse CCT tokenizer. The bottom panel details the convolutional block, composed of multiple Conv1D layers with kernel sizes ranging from 3 to 15, followed by depthwise convolutions and batch normalization. This block was developed and used by three of the authors also in ref.~\cite{mowlaei2025stici}.}
        \label{fig:S2}
\end{figure}

\begin{figure}[t]
        \centering
        \includegraphics[width=7in]{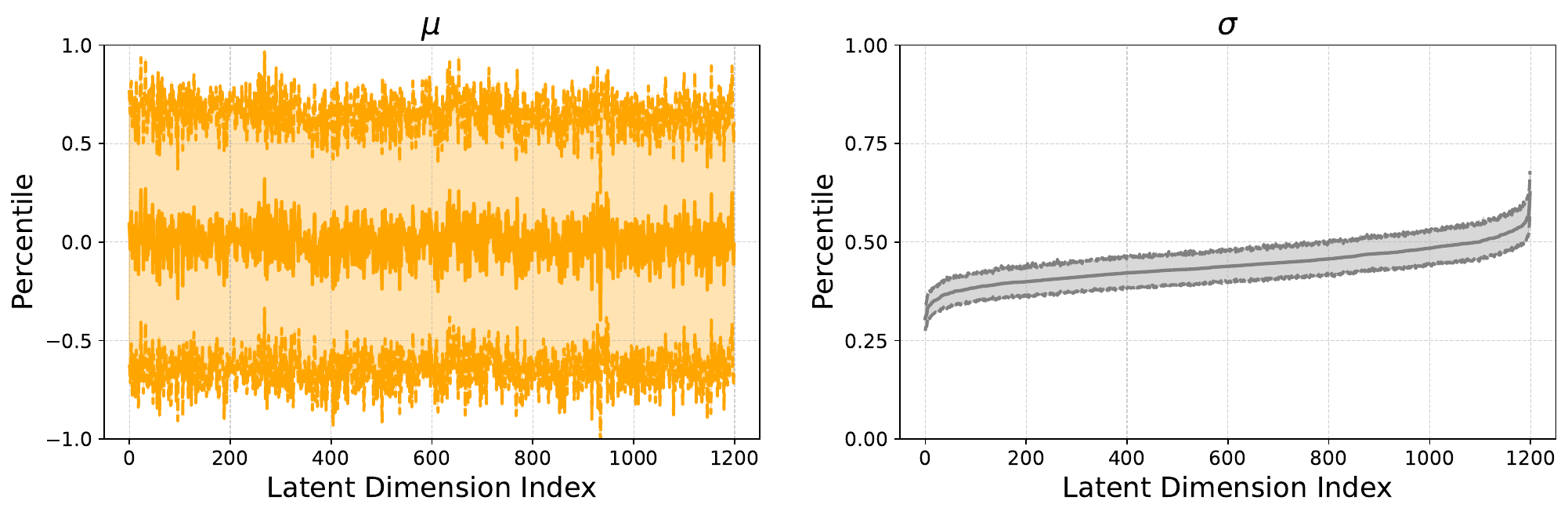}
        \caption{Percentiles of the latent variables across the 1200-dimensional latent space. The plot on the left shows the latent means $\mu$, with the 25th, 50th (median), and 75th percentiles indicated. These are centered around zero and bounded within $\pm 1$, confirming stable encoding across dimensions. The plot on the right displays the latent standard deviations $\sigma$, also reported through the 25th, 50th, and 75th percentiles, with most values lying between 0.3 and 0.6. The latent dimension index is not ordered \emph{a priori}; for visualization purposes, the order was fixed in both plots to enforce a monotonically increasing $\sigma$.}
        \label{fig:S3}
\end{figure}

\begin{figure}[t]
        \centering
        \includegraphics[width=3.5in]{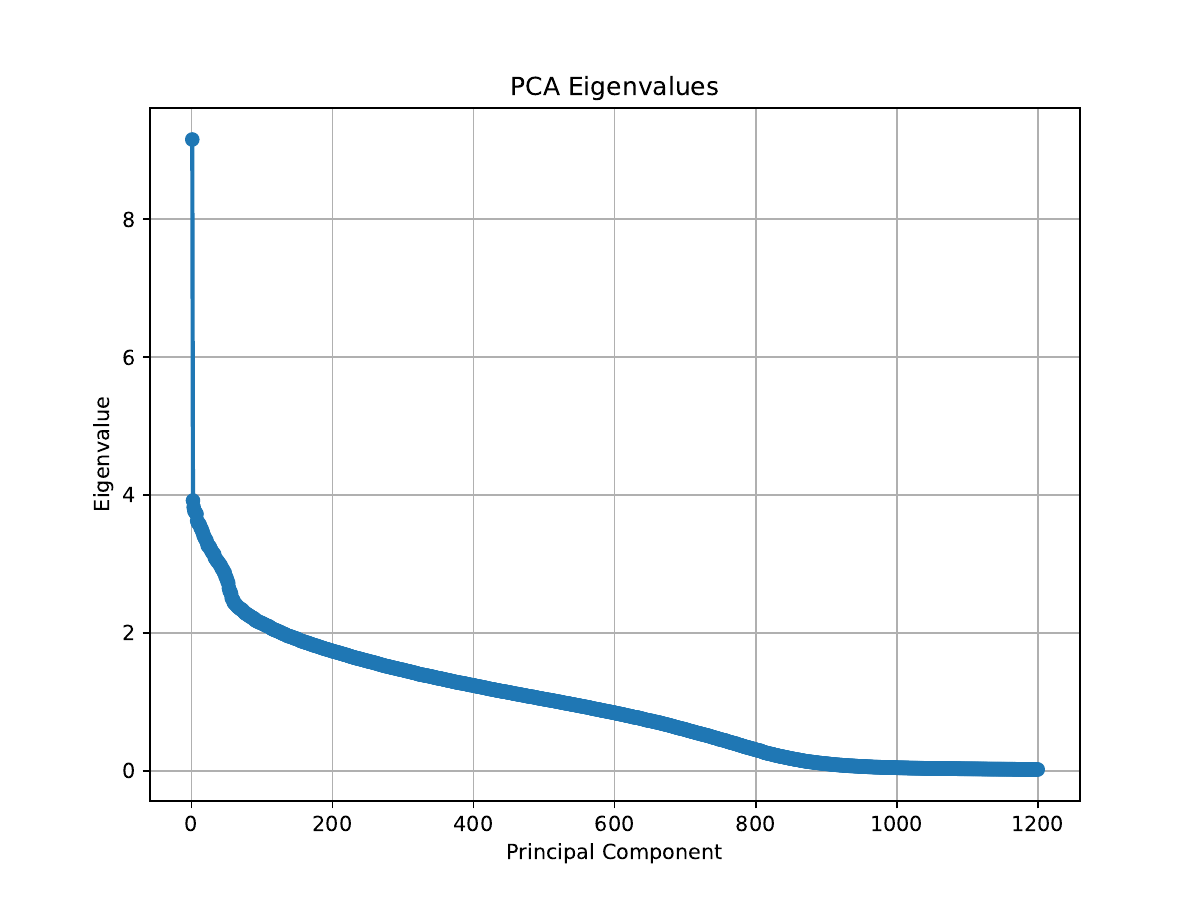}
        \caption{Eigenvalue spectrum from principal component analysis (PCA) of the 1200-dimensional latent space. An initial drop is followed by a gradual decay, indicating that most of the variance is concentrated in the first few hundred components, while higher-order components contribute progressively less. At 1200, the contribution of eigenvalues is negligible, suggesting that the dimension of the space is large enough to encode all the information contained in the training set.}
        \label{fig:S4}
\end{figure}

\begin{figure}[t]
        \centering
        \includegraphics[width=5in]{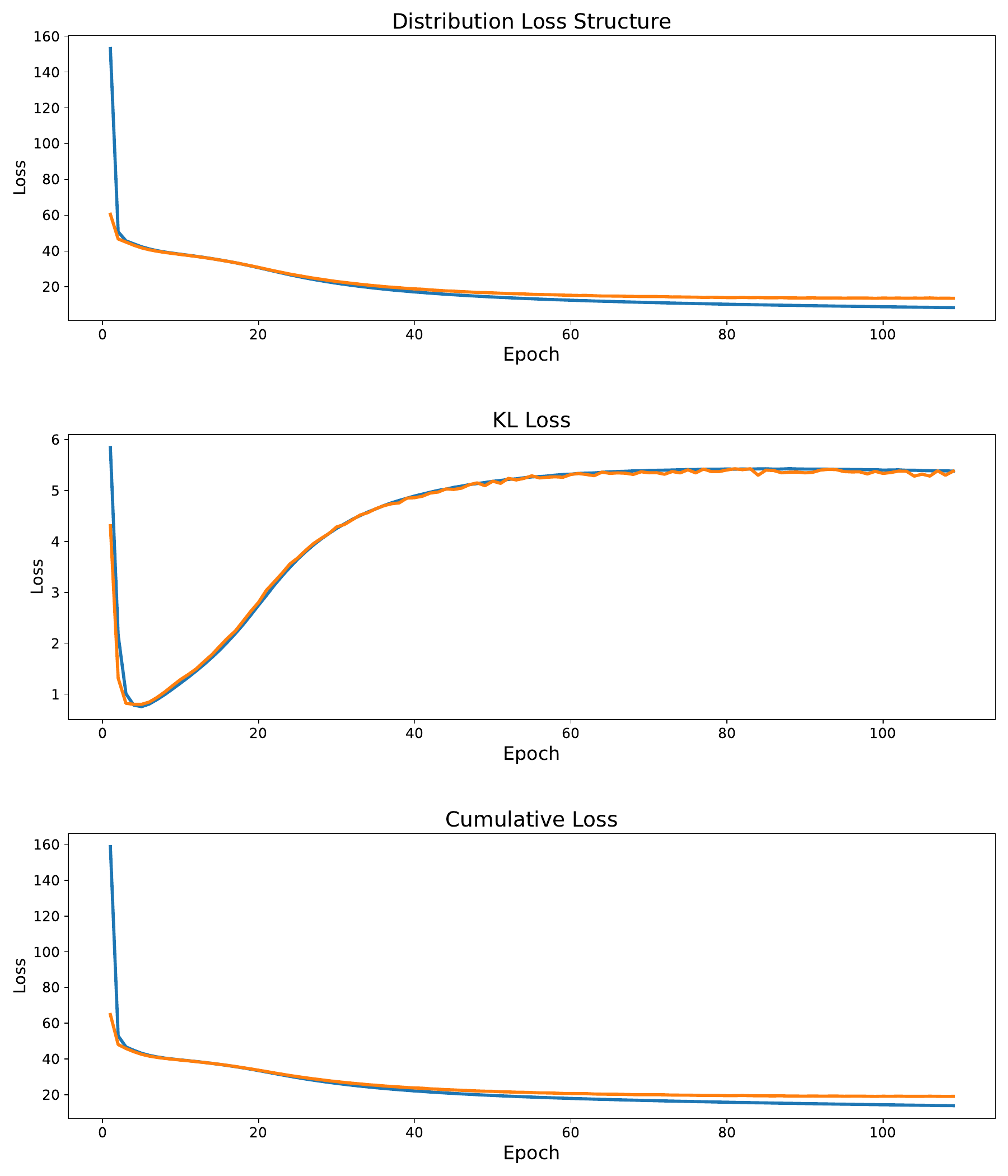}
        \caption{Evolution of the losses over the $120$ training epochs. Top plot shows that the distribution loss, corresponding to the reconstruction error of the input distance matrices, decreases steadily and stabilizes after $\sim$100 epochs. Middle plot shows that the Kullback–Leibler (KL) divergence loss initially decreases sharply, then rises and saturates at a stable value, reflecting regularization of the latent space. The bottom plot shows the cumulative loss, combining reconstruction and KL terms. Training set losses are shown in blue, and validation set losses in orange.}
        \label{fig:S5}
\end{figure}

\begin{figure}[t]
        \centering
        \includegraphics[width=7in]{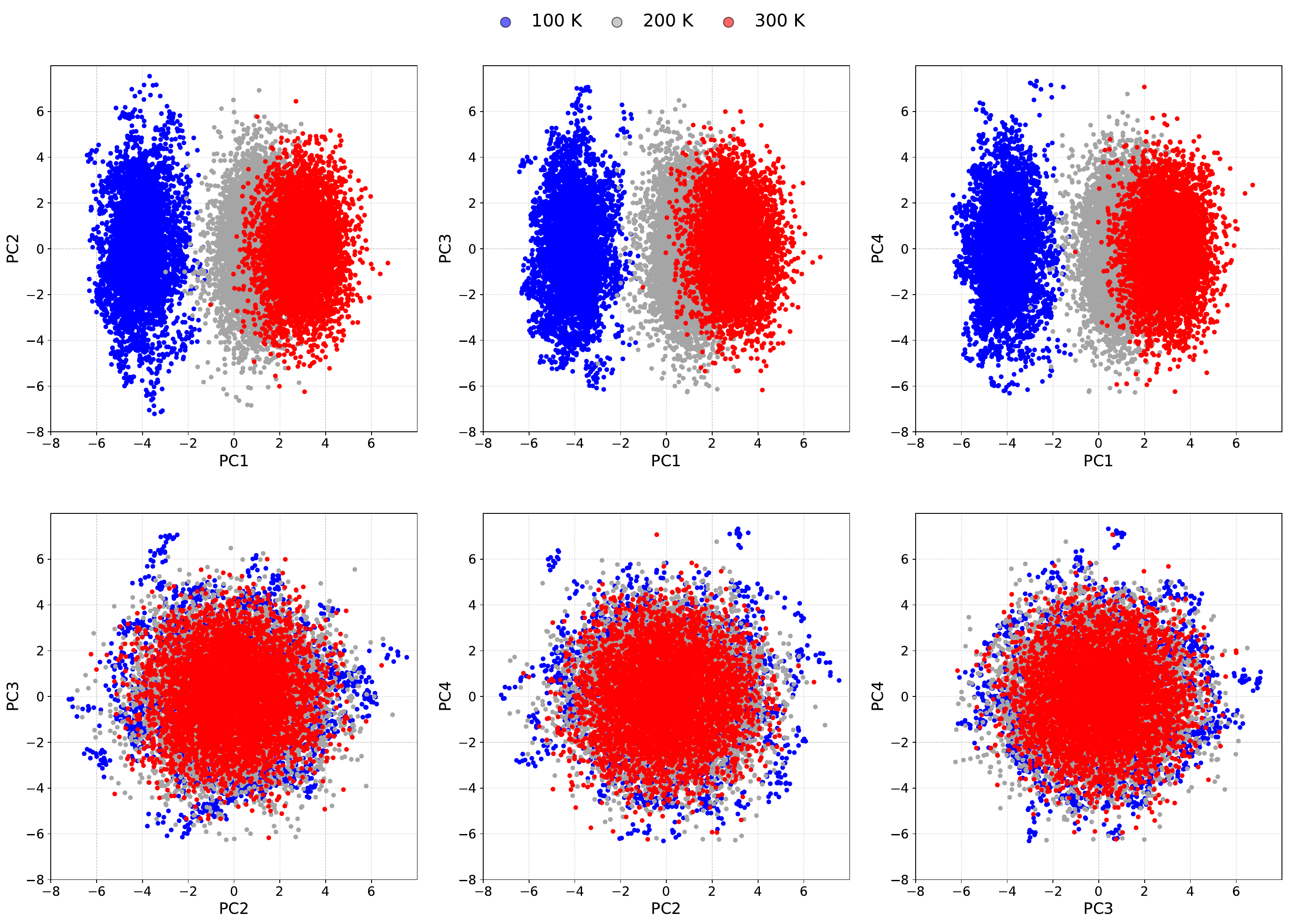}
        \caption{Principal component analysis (PCA) of the 1200-dimensional latent space, projected onto pairs of the first four components. Points are colored by input temperature: $T = 100$ K (blue), $200$ K (gray), and $300$ K (red). Separation of the latent space by temperature is most evident along PC1, while higher-order components (PC2–PC4) exhibit increasing overlap.}
        \label{fig:S6}
\end{figure}

\begin{figure}[t]
        \centering
        \includegraphics[width=7in]{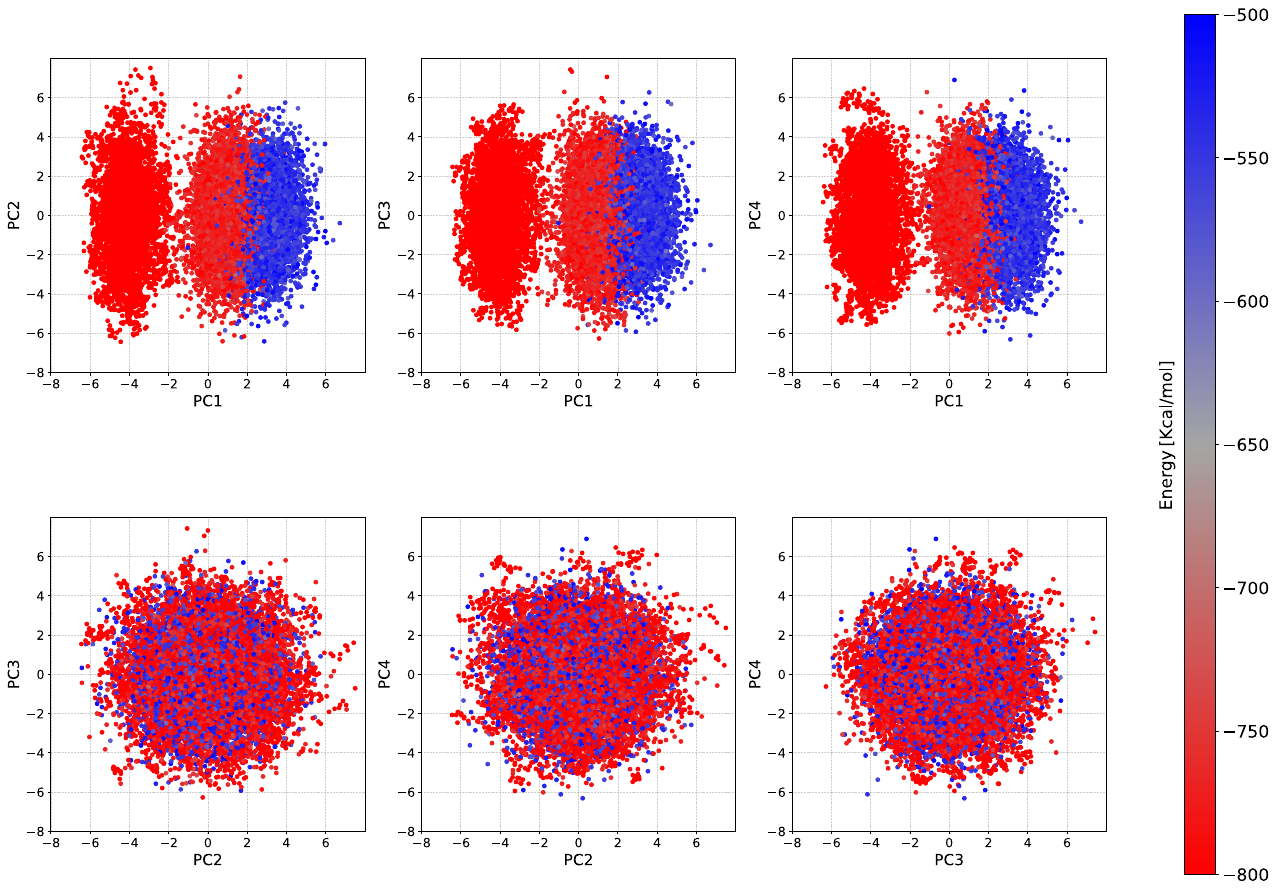}
        \caption{Principal component analysis (PCA) of the 1200-dimensional latent space, projected onto pairs of the first four components. Points are colored by the potential energy of the corresponding polymer configuration, with values ranging from $-800$ Kcal/mol (red) to $-500$ Kcal/mol (blue). Energy correlates strongly with PC1, where low-energy states cluster separately from high-energy ones, while higher-order components (PC2–PC4) show increasing overlap.}
        \label{fig:S7}
\end{figure}

\begin{figure}[t]
        \centering
        \includegraphics[width=7in]{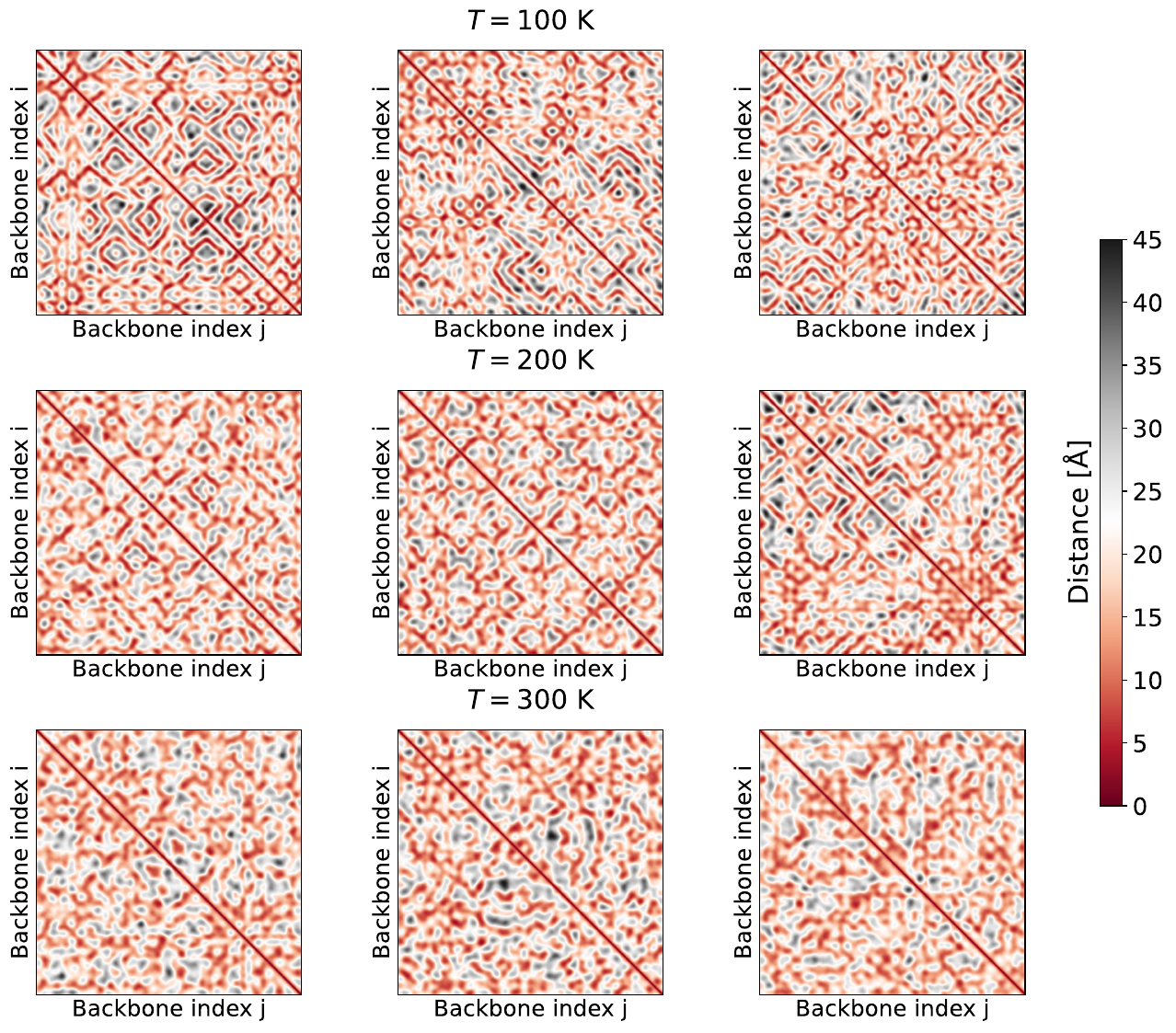}
        \caption{Examples of reconstructed distance matrices of polyethylene chains at $T = 100$, $200$, and $300$ K, corresponding to inputs shown in Fig. S1.}
        \label{fig:S8}
\end{figure}

\begin{figure}[t]
        \centering
        \includegraphics[width=7in]{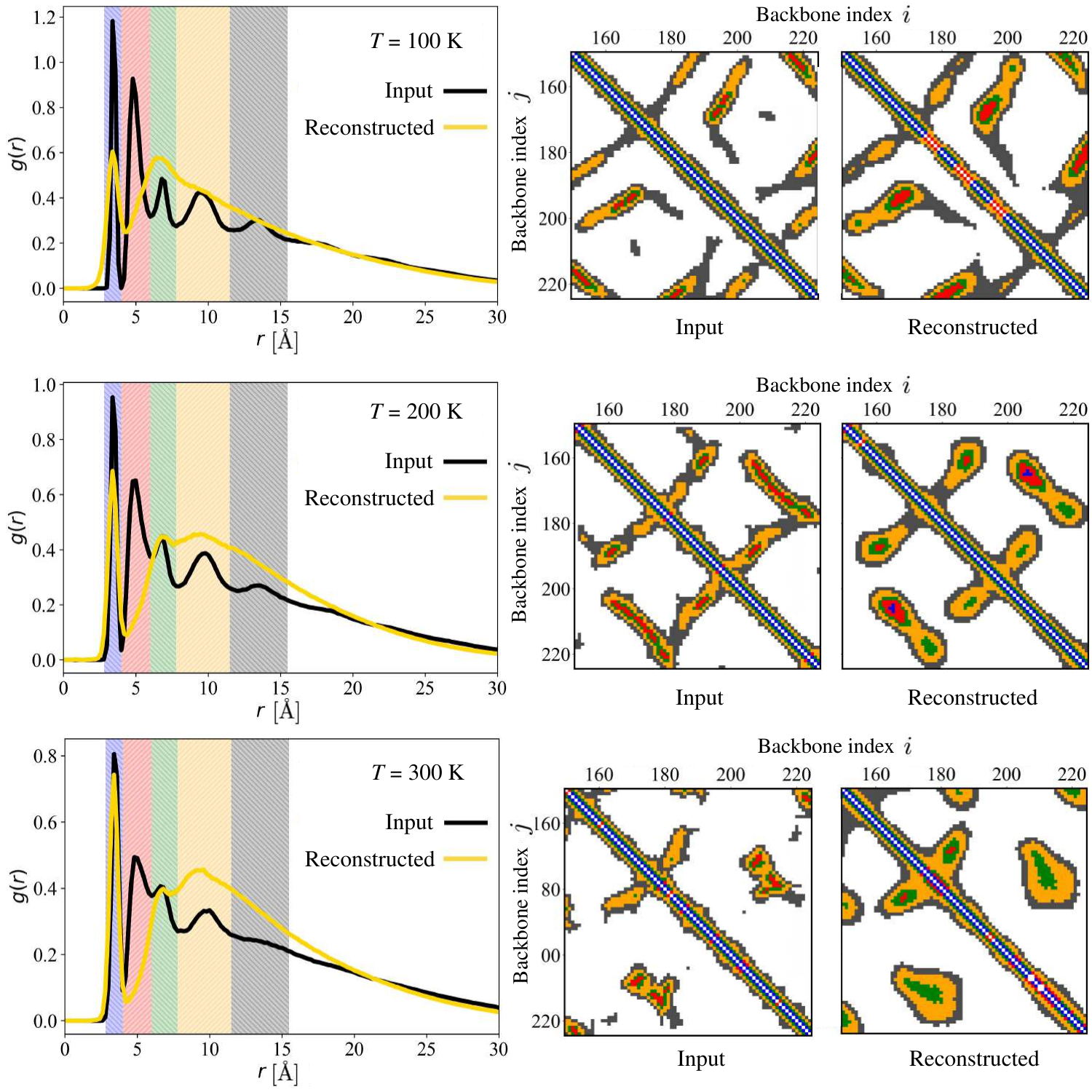}
        \caption{Reconstruction of structural patterns from input distance matrices at $T = 100$, $200$, and $300$ K. On the left column,  the radial distribution functions $g(r)$ of input (black) and reconstructed (yellow) matrices are shown. The main peaks of the input distributions are highlighted by shaded regions and are stable across temperatures.  On the right column, example submatrices ($70 \times 70$) extracted along the main diagonal of input (left) and corresponding reconstructed (right) distance matrices, with elements colored according to the $g(r)$ peaks.}
        \label{fig:S9}
\end{figure}

\begin{figure}[t]
        \centering
        \includegraphics[width=7in]{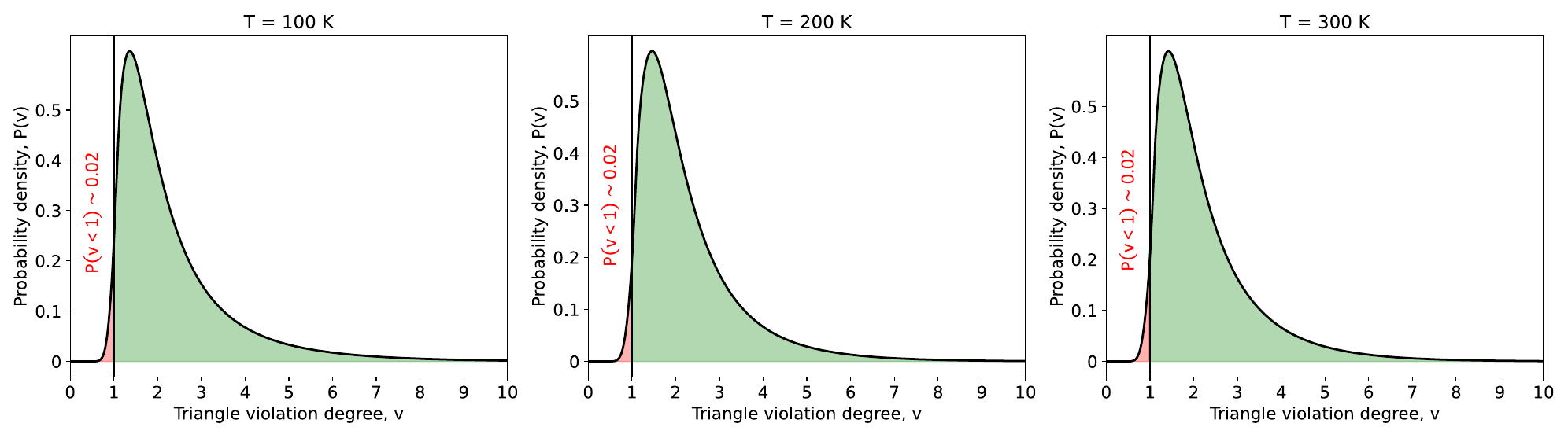}
        \caption{Probability distributions of the triangle inequality violation degree $v$ for all reconstructed distance matrices at $T = 100$, $200$, and $300$ K. For valid distances, the condition $v \geq 1$ must hold. Despite not being explicitly enforced during training, violations with $v < 1$ occur in only $\sim 2\%$ of cases at all temperatures.}
        \label{fig:S10}
\end{figure}

\begin{figure}[t]
        \centering
        \includegraphics[width=7in]{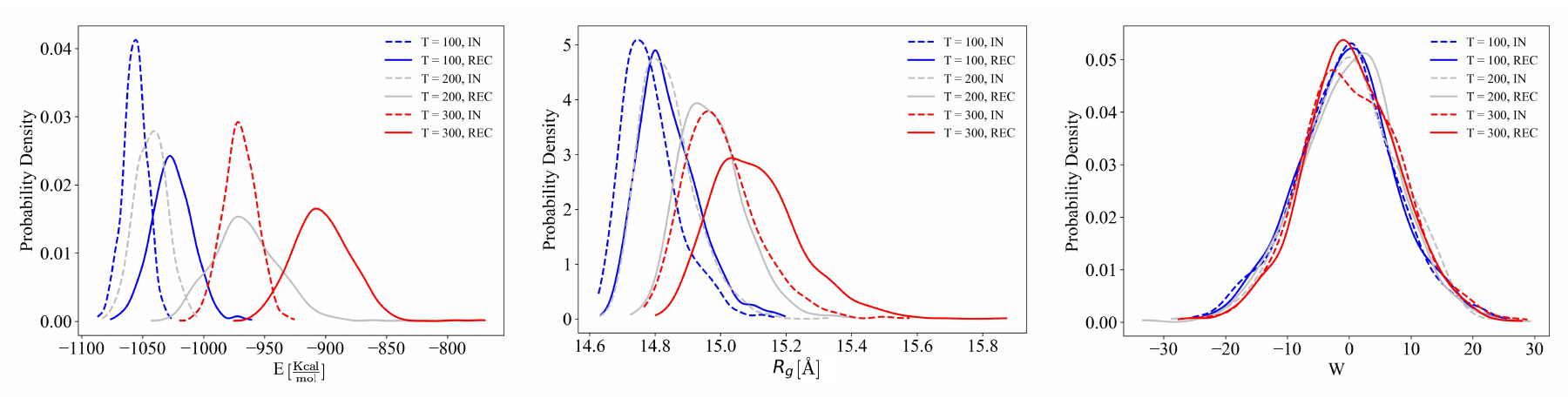}
        \caption{Probability distributions of physical observables for input (dashed lines) and reconstructed (solid lines) samples at $T = 100$ K (blue), $200$ K (gray), and $300$ K (red). (Left) Potential energy $E$, (middle) radius of gyration $R_\mathrm{g}$, and (right) writhe $W$. Reconstructed distributions of $W$ follow the input ones for all the explored temperatures, while $E$ and $R_\mathrm{g}$ are generally shifted upward. All distributions are smoothed using a Gaussian kernel density estimate (KDE) for visual clarity.}
        \label{fig:S11}
\end{figure}

\begin{figure}[t]
        \centering
        \includegraphics[width=3.5in]{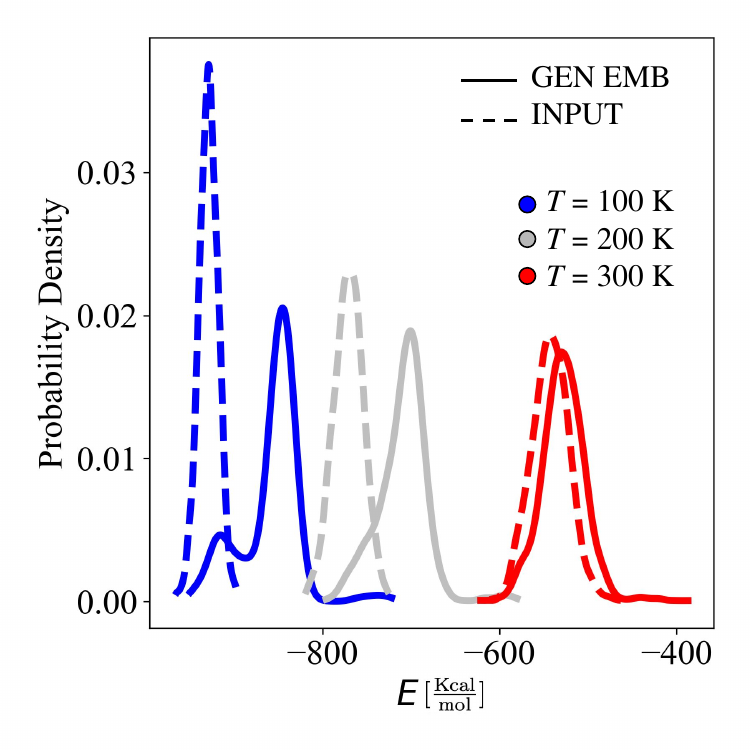}
        \caption{Probability distributions of the potential energy $E$ for input (dashed lines) and generated embedded (solid lines) configurations at $T = 100$ K (blue), $200$ K (gray), and $300$ K (red). All distributions are smoothed using a Gaussian kernel density estimate (KDE) for visual clarity.}
        \label{fig:S12}
\end{figure}

\begin{figure}[t]
        \centering
        \includegraphics[width=7in]{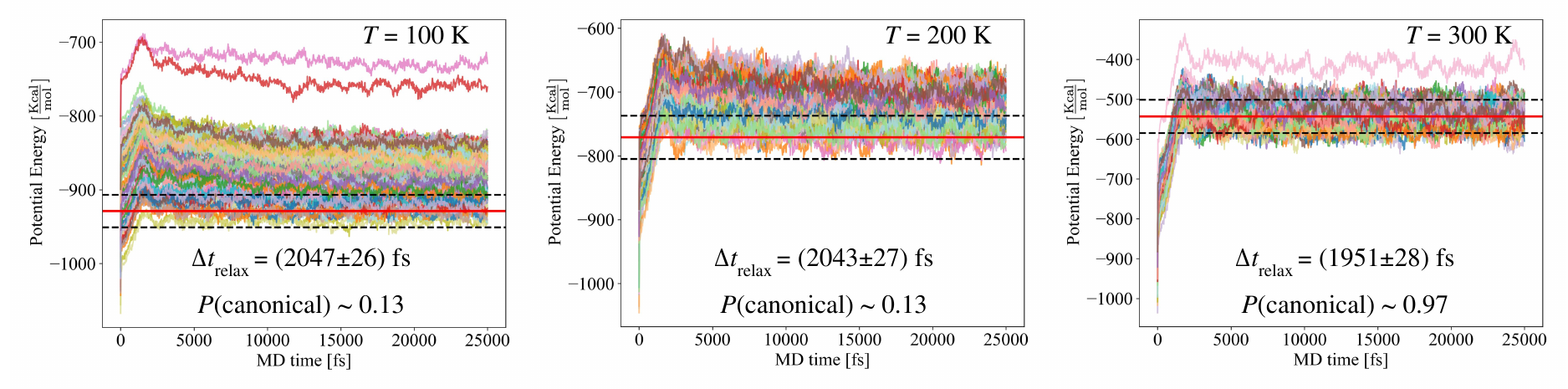}
        \caption{Potential energy trajectories during MD relaxation of $100$ generated samples at $T = 100$, $200$, and $300$ K. Convergence is reached for times $\sim 2$ ps, with average relaxation times $\Delta t_\mathrm{relax} = (2047 \pm 26)$ fs at 100 K, $(2043 \pm 27)$ fs at 200 K, and $(1951 \pm 28)$ fs at 300 K. The canonical ensemble mean computed from the training set, $\langle E \rangle$ (solid red line) and standard deviation $\sigma$ (dashed black lines) are indicated. The fraction of relaxed configurations within $\langle E \rangle \pm 2\sigma$, $P(\mathrm{canonical})$, is $\sim 0.13$ at $100$ and 200 K, but rises to $\sim 0.97$ at $300$ K. The stationary was quantified using an Augmented Dickey–Fuller test with 5\% confidence.}
        \label{fig:S13}
\end{figure}

\begin{figure}[t]
        \centering
        \includegraphics[width=7in]{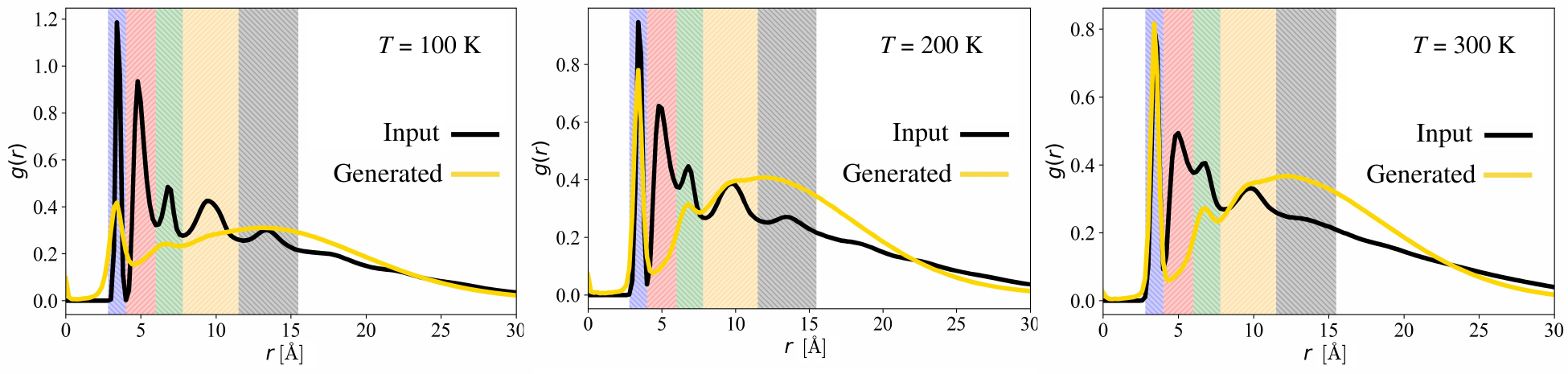}
        \caption{Radial distribution functions $g(r)$ of input (black) and generated (yellow) distance matrices obtained directly from the decoder at $T = 100$, $200$, and $300$ K. While the generated outputs reproduce some of the qualitative features of the input distributions, they tend to underestimate intermediate-range peaks and overestimate long-range density.}
        \label{fig:S14}
\end{figure}

\begin{figure}[t]
        \centering
        \includegraphics[width=7in]{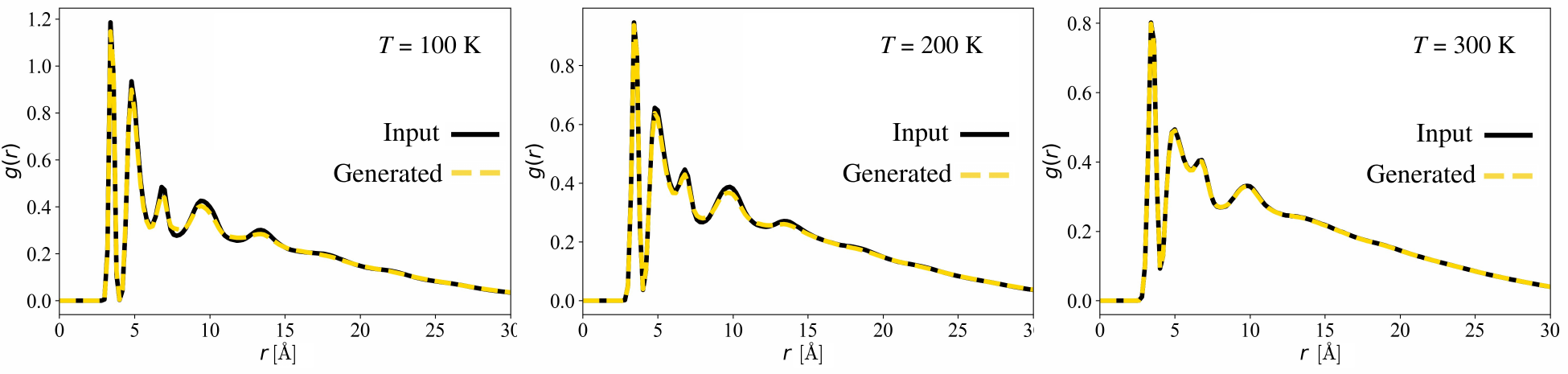}
        \caption{Radial distribution functions $g(r)$ of input (black) and generated embedded configurations (dashed yellow) at $T = 100$, $200$, and $300$ K. After the embedding and relaxation procedure, the generated samples reproduce the canonical input radial distributions with good agreement across all temperatures.}
        \label{fig:S15}
\end{figure}

\begin{figure}[t]
        \centering
        \includegraphics[width=7in]{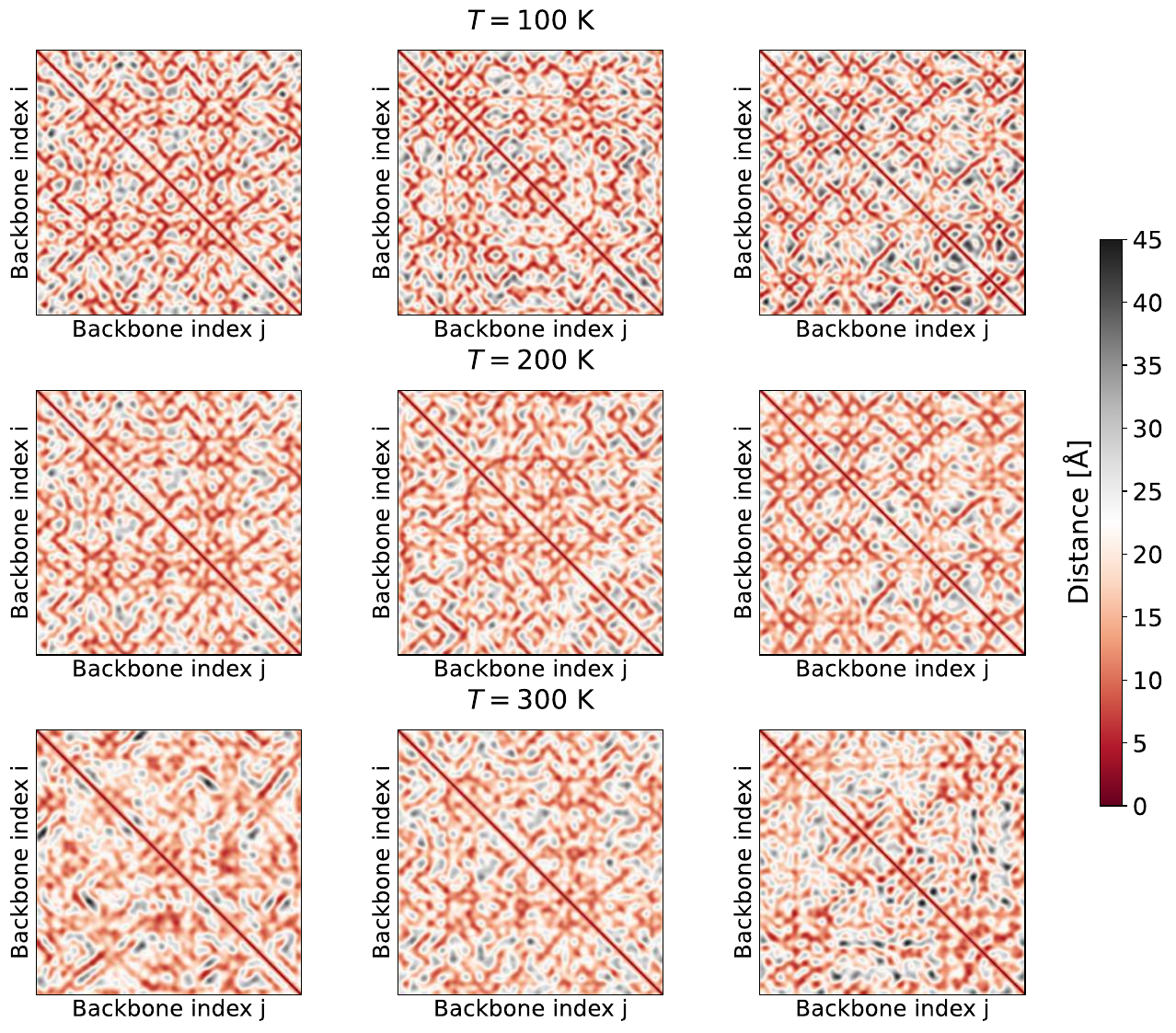}
        \caption{Examples of generated distance matrices at $T = 100$, $200$, and $300$ K, obtained directly from the decoder. These examples illustrate physically consistent outputs, which will have the correct energy value after the embedding procedure.}
        \label{fig:S16}
\end{figure}

\begin{figure}[t]
        \centering
        \includegraphics[width=7in]{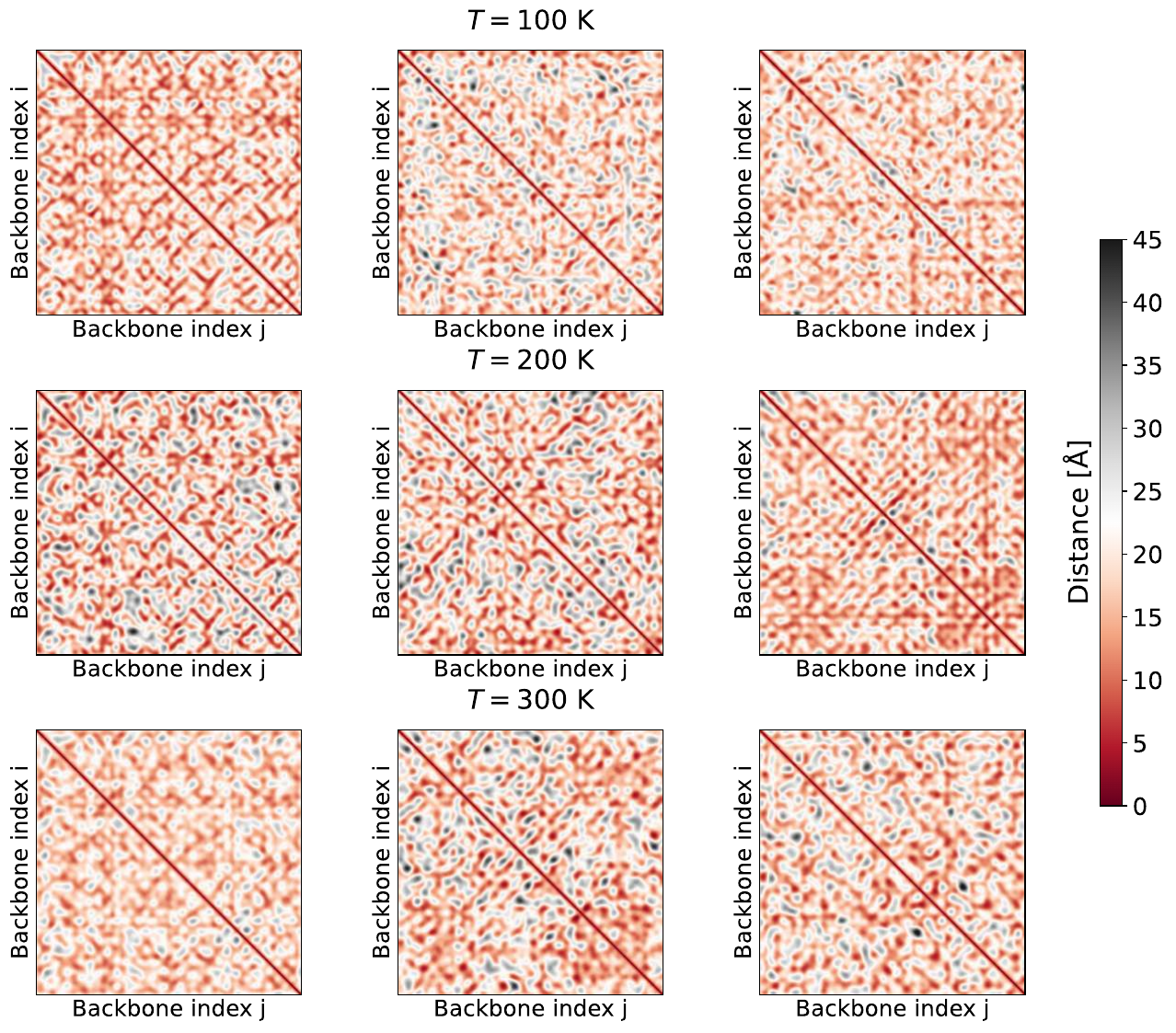}
        \caption{Examples of generated distance matrices at $T = 100$, $200$, and $300$ K, obtained directly from the decoder. These examples correspond to hallucination outputs, which will be discarded after the embedding procedure because of their non-physical energy.}
        \label{fig:S17}
\end{figure}

\begin{figure}[t]
        \centering
        \includegraphics[width=7in]{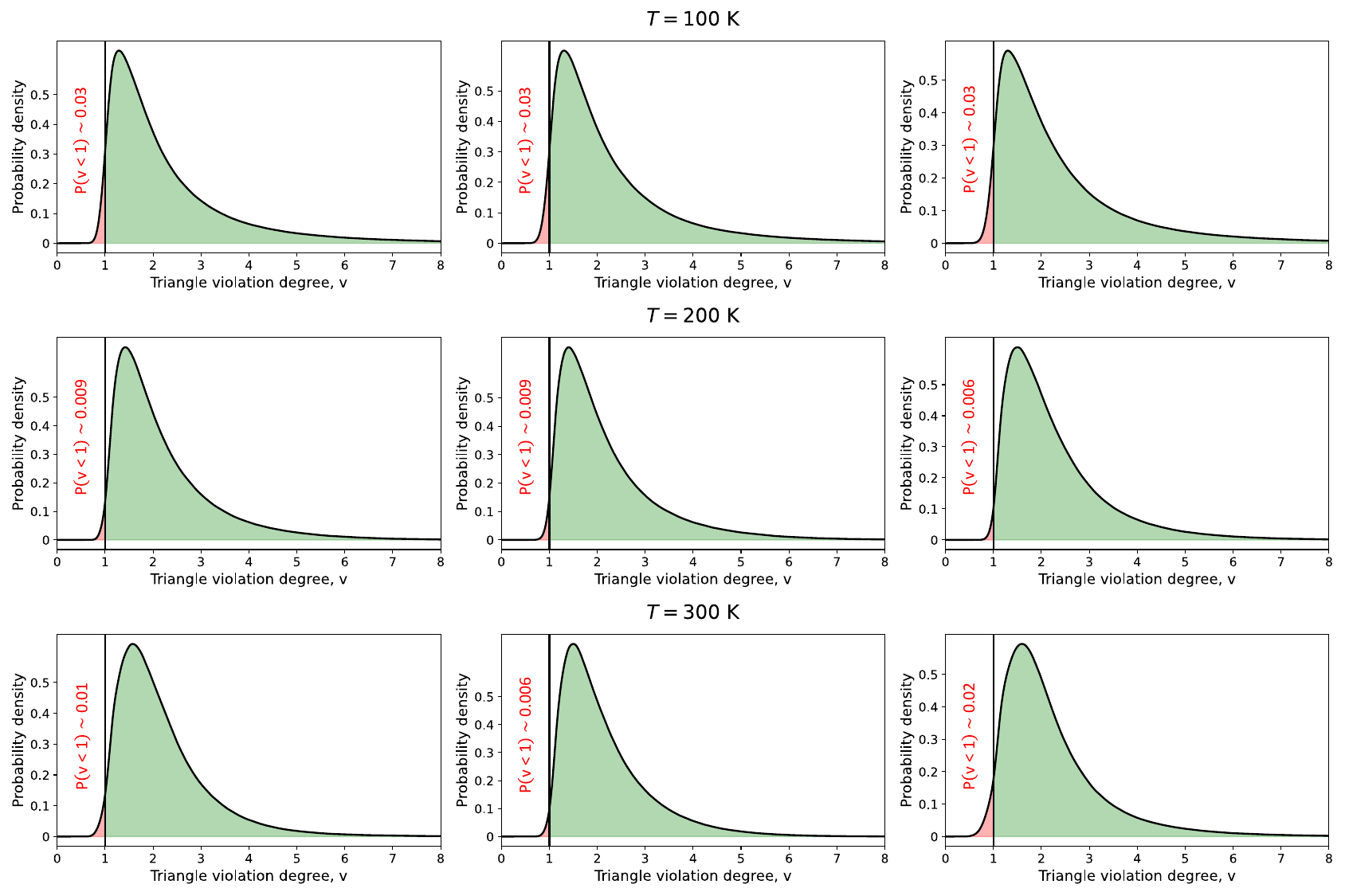}
        \caption{Probability distributions of the triangle inequality violation degree $v$ for the same physically consistent generated matrices shown in Fig. S16, at $T = 100$, $200$, and $300$ K. For valid distances, $v \geq 1$ must hold. Despite not being explicitly enforced, violations with $v < 1$ occur in only $\sim 3\%$  of cases at $T = 100$ K, less than $\sim 1\%$ at $T = 200$ K and less than $\sim 2\%$ at $T = 300$ K.}
        \label{fig:S18}
\end{figure}

\begin{figure}[t]
        \centering
        \includegraphics[width=7in]{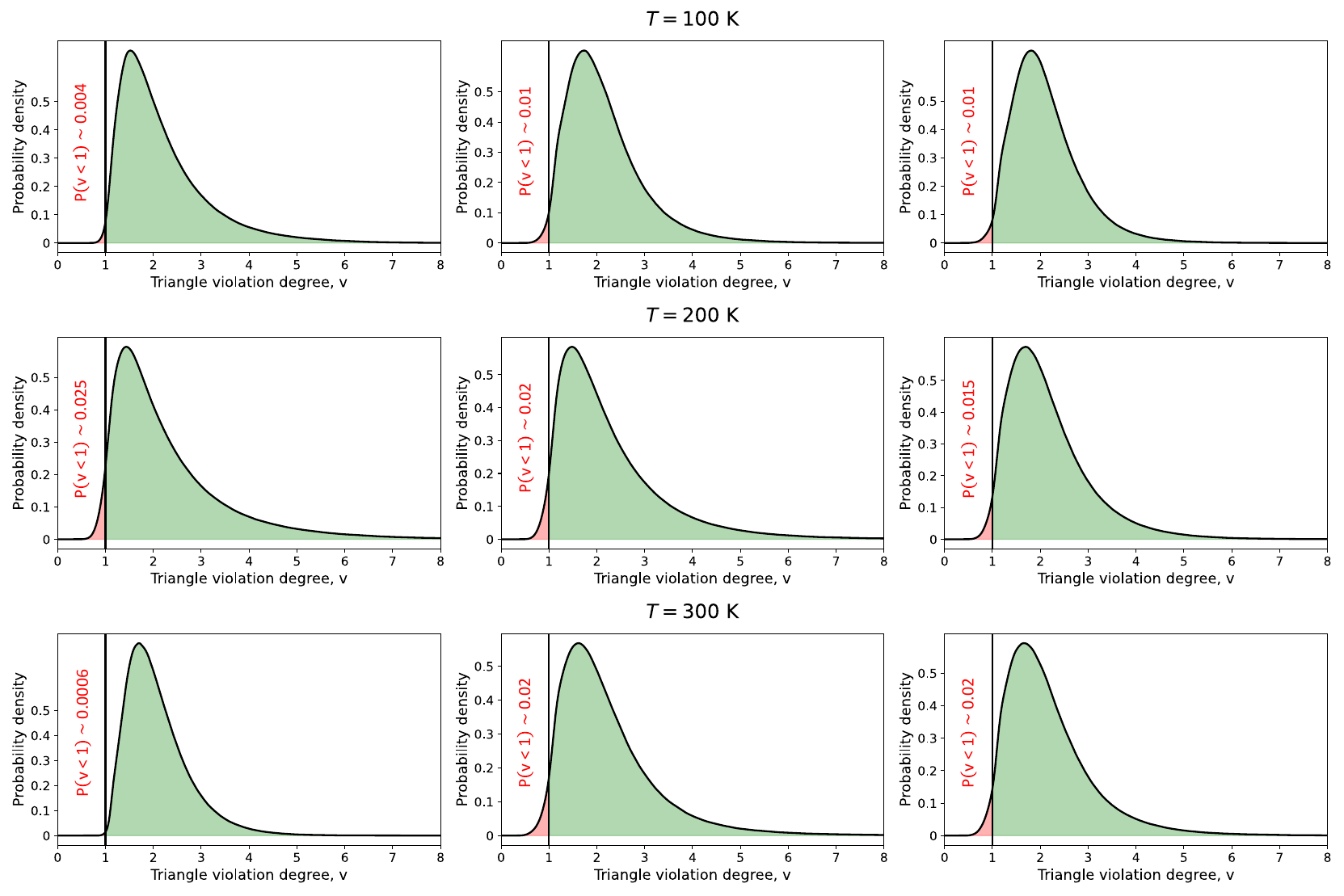}
        \caption{Probability distributions of the triangle inequality violation degree $v$ for the same nonphysical generated matrices shown in Fig. S17, at $T = 100$, $200$, and $300$ K. For the these samples, violations with $v < 1$ occur in less than $\sim 1\%$  of cases at $T = 100$ K, $\sim 2\%$ or more at $T = 200$ K and less than $\sim 2\%$ at $T = 300$ K.}
        \label{fig:S19}
\end{figure}

\begin{figure}[t]
        \centering
        \includegraphics[width=7in]{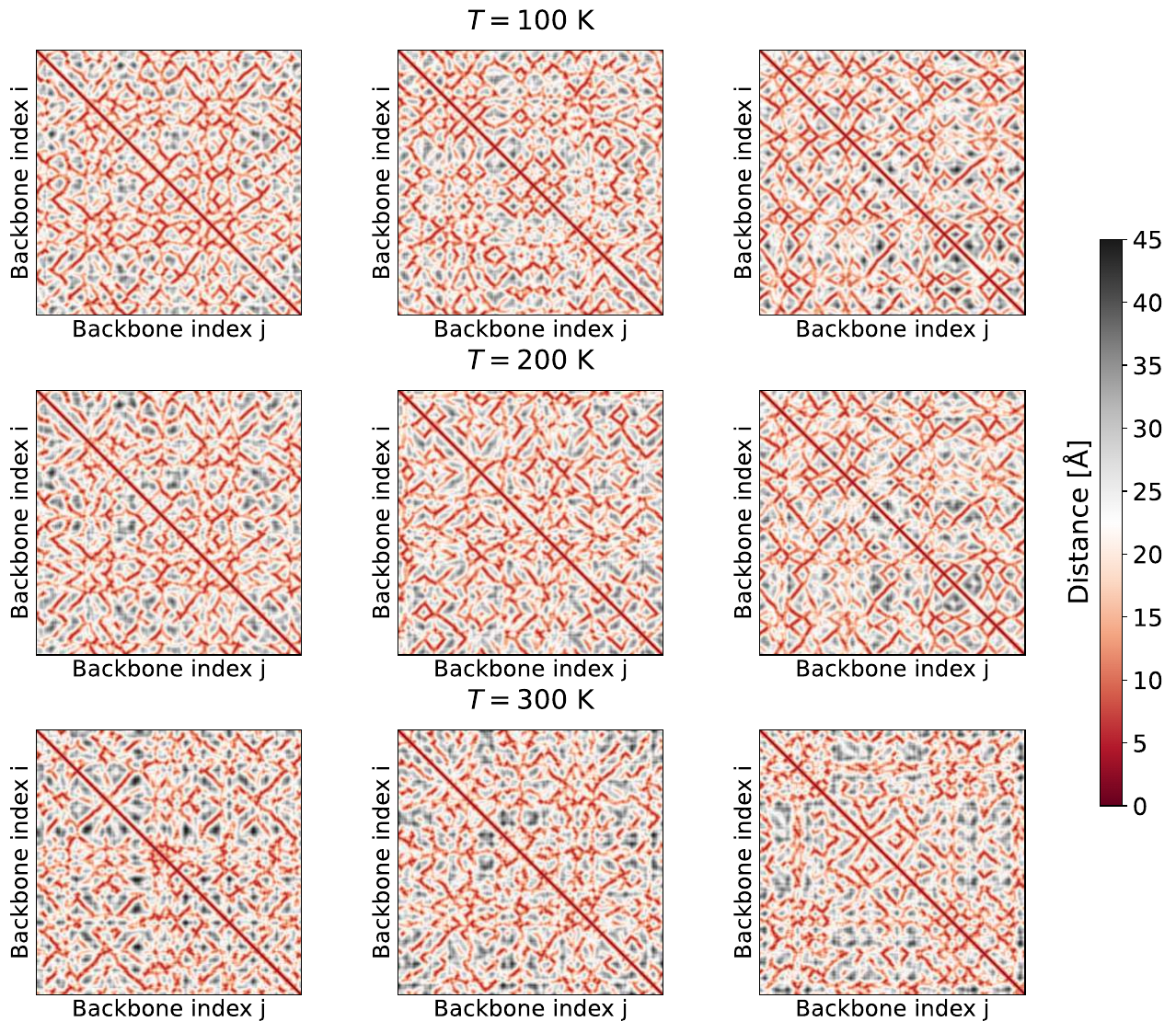}
        \caption{Examples of generated distance matrices at $T = 100$, $200$, and $300$ K, obtained after embedding and relaxation of the decoder outputs shown in Fig. S16.}
        \label{fig:S20}
\end{figure}

\begin{figure}[t]
        \centering
        \includegraphics[width=7in]{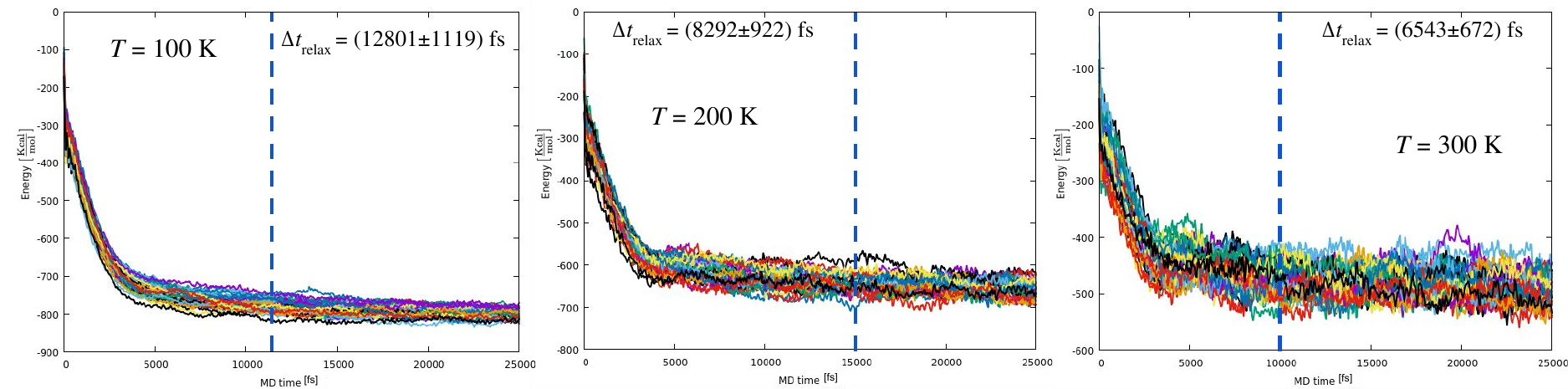}
        \caption{Potential energy trajectories during MD relaxation of input samples from swollen self-avoiding random walks to globular the phase at $T = 100$, $200$, and $300$ K. Convergence is reached for times $\sim 15$ ps, with average relaxation times $\Delta t_\mathrm{relax} = (12801 \pm 1119)$ fs at 100 K, $(8292 \pm 922)$ fs at 200 K, and $(6543 \pm 672)$ fs at 300 K. The stationary was quantified using an Augmented Dickey–Fuller test with 5\% confidence.}
        \label{fig:S21}
\end{figure}

\begin{figure}[t]
        \centering
        \includegraphics[width=7in]{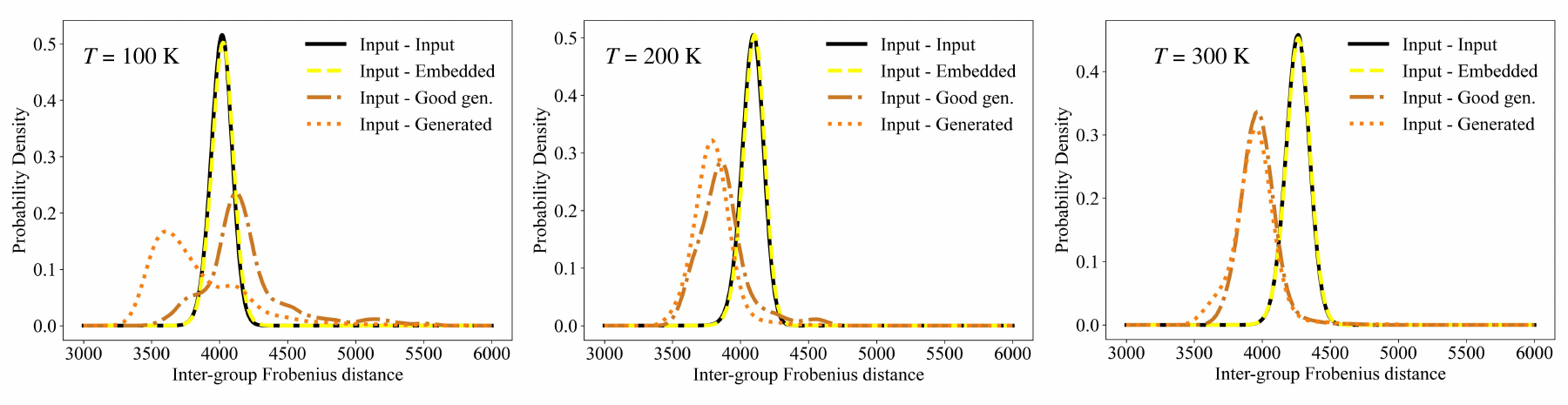}
        \caption{Novelty analysis of generated samples using all Frobenius distances in real space. Probability distributions of all inter-group Frobenius distances at $T = 100$, $200$, and $300$ K. Comparisons are shown between input-input (black), input-generated raw decoder outputs (orange), input-energy filtered outputs (brown), and input-embedded configurations (yellow) pairs. As in Fig.6a in the main text, generated outputs become statistically indistinguishable from inputs only after the embedding step. All distributions are smoothed using a Gaussian KDE for visual clarity.}
        \label{fig:S22}
\end{figure}

\begin{figure}[t]
        \centering
        \includegraphics[width=7in]{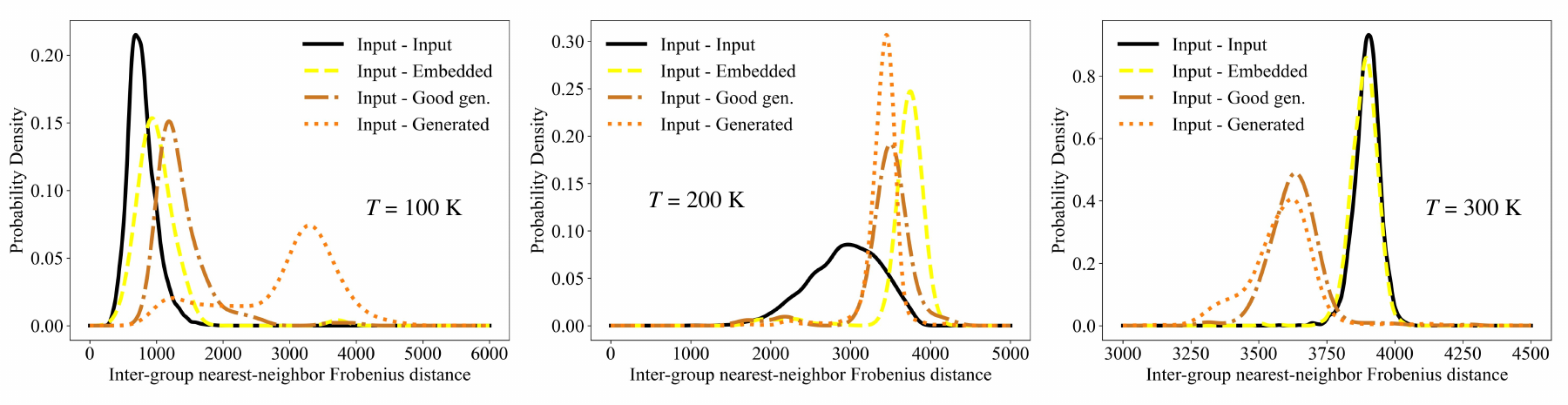}
        \caption{Novelty analysis of generated samples using nearest-neighbor Frobenius distances in real space. Probability distributions of only inter-group nearest-neighbor Frobenius distances at $T = 100$, $200$, and $300$ K for input-input (black), input-generated raw decoder outputs (orange), input-energy filtered outputs (brown), and input-embedded configurations (yellow) pairs. As in Fig.6b in the main text, the input–embedded distributions are shifted relative to input–input for all temperatures, indicating that generated samples correspond to novel configurations beyond the training data. All distributions are smoothed using a Gaussian KDE for visual clarity.}
        \label{fig:S23}
\end{figure}

\begin{figure}[t]
        \centering
        \includegraphics[width=7in]{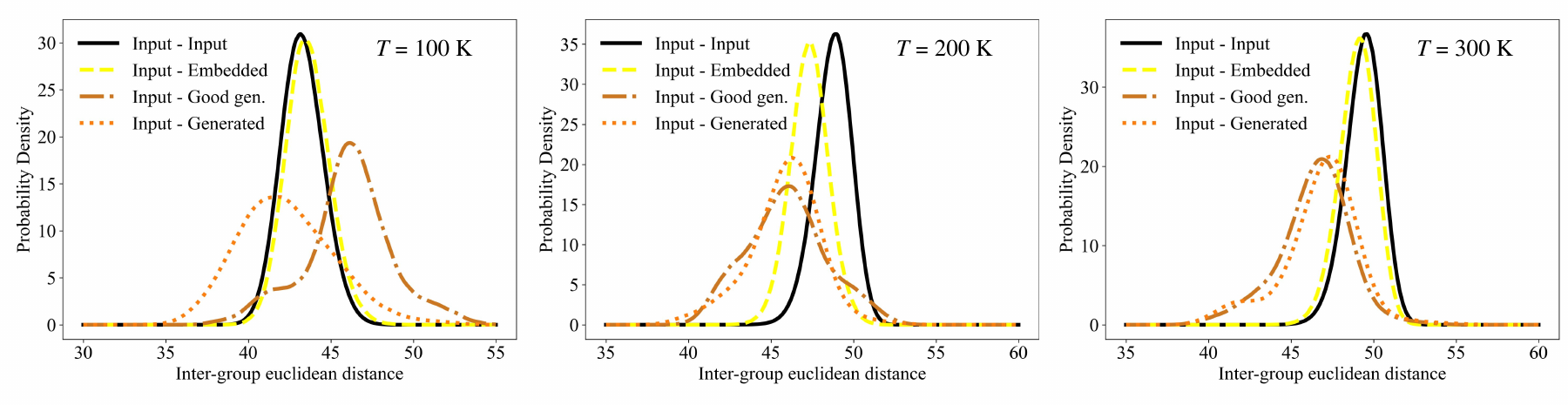}
        \caption{Novelty analysis of generated samples in latent space using all Euclidean distances. Probability distributions of all inter-group Euclidean distances at $T = 100$, $200$, and $300$ K, computed using the mean ($\mu$) of each latent vectors. For input samples, $\mu$ values were taken directly from the encoder; for generated samples, matrices were re-embedded in the latent space using the encoder and their corresponding latent $\mu$ vectors extracted. The qualitative behavior mirrors that observed in real space (Fig. S21), suggesting that the latent representation based on $\mu$ is approximately distance-preserving up to a scaling factor. Comparisons are shown between input-input (black), input-generated raw decoder outputs (orange), input-energy filtered outputs (brown), and input-embedded configurations (yellow) pairs. All distributions are smoothed using a Gaussian KDE for visual clarity.}
        \label{fig:S24}
\end{figure}

\begin{figure}[t]
        \centering
        \includegraphics[width=7in]{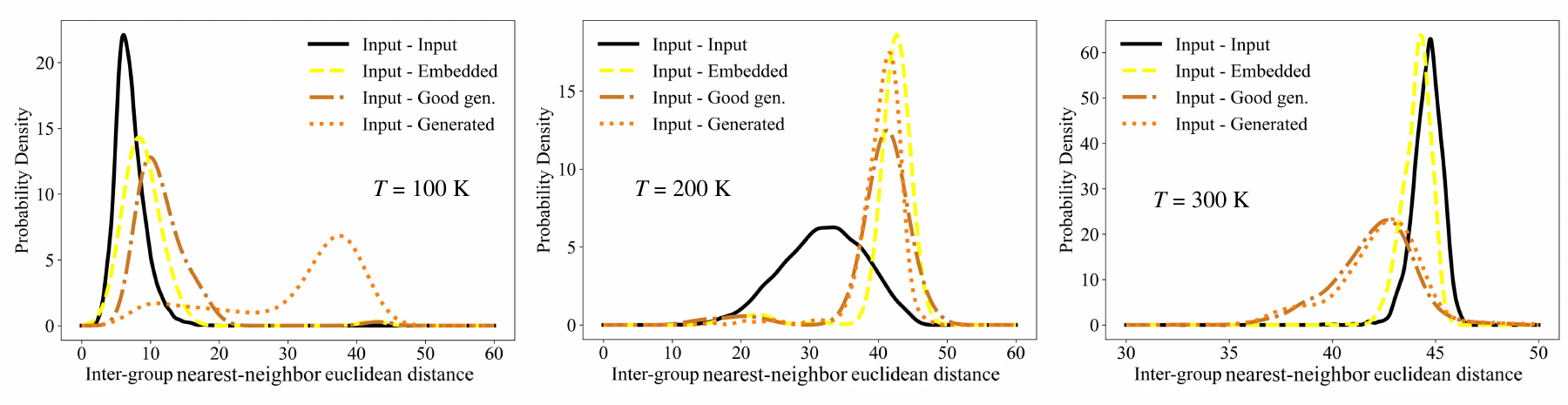}
        \caption{Novelty analysis of generated samples in latent space using nearest-neighbor Euclidean distances. Probability distributions of only inter-group nearest-neighbor Euclidean distances at $T = 100$, $200$, and $300$ K, computed using the mean ($\mu$) of each latent vectors. Comparisons are shown between input-input (black), input-generated raw decoder outputs (orange), input-energy filtered outputs (brown), and input-embedded configurations (yellow) pairs. For input samples, $\mu$ values were taken directly from the encoder; for generated samples, matrices were re-embedded in the latent space using the encoder and their corresponding latent $\mu$ vectors extracted. As in Fig. S22, the input–embedded distributions are shifted to larger values relative to input–input for all temperatures. All distributions are smoothed using a Gaussian KDE for visual clarity.}
        \label{fig:S25}
\end{figure}

\clearpage

\bibliography{bibliography}